\documentclass[iop]{emulateapj}
\usepackage{amsmath,amssymb}
\usepackage{graphicx}

\newcommand{\lyman}{Ly$\alpha~$}

\newcommand{\asec}{$^{\prime\prime}~$}

\newcommand{\iion}[2]{\hbox{#1\,{\sc #2}}}

\slugcomment{}

\shorttitle{Automated and Objective Detection of Emission Lines}
\shortauthors{Hong et al.}

\begin{document}


\title{On the Automated and Objective Detection of Emission Lines in Faint-Object Spectroscopy }

\author{ 
Sungryong Hong\altaffilmark{1}, 
Arjun Dey\altaffilmark{1,}\altaffilmark{2}, and  
Moire K. M. Prescott\altaffilmark{3}
}

\altaffiltext{1}{National Optical Astronomy Observatory,  
Tucson, AZ 01003, USA}
\altaffiltext{2}{Radcliffe Fellow, Radcliffe Institute of Advanced Study, 
Byerly Hall, Harvard University, 10 Garden Street, Cambridge, MA 02138}
\altaffiltext{3}{Dark Cosmology Centre, Niels Bohr Institute, University of Copenhagen, 
Juliane Maries Vej 30, 2100 Copenhagen {\O}, Denmark}

\begin{abstract}
Modern spectroscopic surveys produce large spectroscopic databases, generally with sizes well beyond the scope of manual investigation.  
The need arises, therefore, for an automated line detection method with objective indicators for detection significance. 
In this paper, we present an automated and objective method for emission line detection in spectroscopic surveys and apply 
this technique to 1574 spectra, obtained with the Hectospec spectrograph on the MMT Observatory (MMTO), to detect Lyman alpha emitters near $z \sim 2.7$. 
The basic idea is to generate on-source (signal plus noise) and off-source (noise only) mock observations 
using Monte Carlo simulations, and calculate completeness and reliability values, $(C, R)$, for each simulated signal. 
By comparing the detections from real data with the Monte Carlo results, we assign the completeness and reliability values to each real detection. 
From 1574 spectra, we obtain 881 raw detections and, by removing low reliability detections, we finalize 649 detections from an automated pipeline. 
Most of high completeness and reliability detections, $(C, R) \approx (1.0, 1.0)$, 
are robust detections when visually inspected; 
the low C and R detections are also marginal on visual inspection. 
This method at detecting faint sources is dependent on the accuracy of the sky subtraction.


\end{abstract}
\keywords{methods: data analysis -- techniques: spectroscopic}

%
%


 \section{Introduction}
Modern spectroscopic surveys produce large and relatively uniform spectroscopic databases 
(e.g. AGES, BOSS, BigBOSS, and SEGUE; Kochanek et al 2012, Dawson et al. 2013, Schlegel et al. 2009 ,Yanny et al. 2009). 
Though many preprocessing reduction pipelines are automated, 
the reduced spectra traditionally have been analyzed visually. 
Particularly with the advent of the Sloan Digital Sky Survey (SDSS; York et al. 2000, Ahn et al. 2012), 
we have now entered an era of 
immense spectroscopic databases which defy manual inspection and necessitate automation. 
For example, Bolton et al. (2012) presented automated pipelines to classify galaxies and measure their redshifts from BOSS spectra, 
and Lee et al. (2008) did the same to measure stellar properties such as radial velocity, effective temperature, and metallicity from SEGUE spectra. 

In this paper, we present an objective and automated method for the detection of line emission in spectroscopic data.  
This method is designed for the detection of single emission lines (e.g., arising from \lyman emission) within a certain spectral range; 
in our application, 4000 -- 5000\AA. Since the continuum is usually undetected in the observed spectra of most \lyman emitting galaxies,  
our method mainly focuses on discerning real astronomical signal from coincident noise features by providing quantitative measurements 
of detection significance. 
Although the main application of our method is to search for faint emission lines on a flat (continuum-free) noise background, 
we can extend our method to more generic spectra, if continuum baselines can be removed properly. 

The outline of our paper is as follows. In \S 2, we describe the basic algorithm  
and present results from Monte Carlo (MC) simulations. 
In \S 3, we apply our technique to 1574 sample spectra obtained using the Hectospec instrument on the MMT. 
We summarize our findings in \S 4.

\section{Method Description}\label{sec:two}
The problem we address in this paper is how to automatically extract faint emission features (called ``detections'') 
and quantify their significance. 
For this purpose, we employ commonly used indicators of detection significance : 
``reliability'' and ``completeness''. ``Reliability'' is the probability that a given detection is real (i.e.,, not resulting from noise). 
``Completeness'' is a measure which quantifies the detectability of underlying signal source of a given flux. 
A completeness equal to 1.0 implies that the feature is detected consistently in all multiple observations of similar depth. 
We describe, in what follows, the mathematical frame work used in this paper 
and then calculate the reliability and completeness  using mock observations from Monte Carlo (MC) simulations.

\subsection{Basic Idea}\label{sec:bi}

In this section, we present our mathematical framework and definitions of completeness and reliability. 
Then, we describe a practical implementation of our method in \S2.2. 
One can focus more on the implementation section, if mathematical rigor is not necessary. 

\subsubsection{Mathematical Definitions}

A signal detection algorithm must quantitatively describe three parts : 
(1) the intrinsic signals, (2) the measured features, and (3) the detection criteria. 

We can parametrize signals using a set of mathematical quantities. 
We call this set of parameters defining the signal as the ``source vector'', $\vec{s}$,
which spans a ``signal space''. Signals are represented as points in this signal space. 
For example, Gaussian signals can be parametrized by 3 numbers, i.e, their centroid, width and height.

\begin{figure}
\centering
\includegraphics[height=2.5 in]{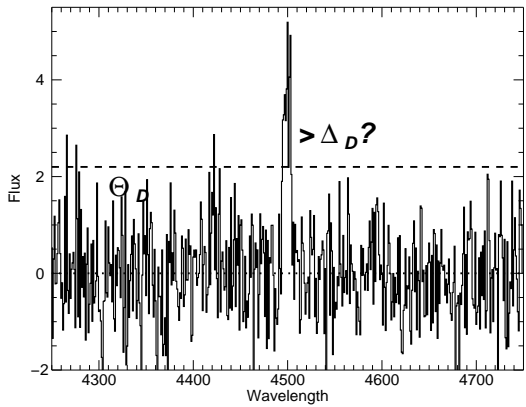}
\caption{A mock spectrum ranging 4250--4750\AA ~with an underlying Gaussian signal, 
consisting of random noise with a Gaussian-profile emission line at $\lambda=4500$\AA. 
We call the three parameters representing the underlying Gaussian profile the ``source vector''. 
The dashed horizontal line represents the threshold line with the height of $\Theta_D$. 
We define ``detection'' if the width of the feature above the threshold line is larger than $\Delta_D$. 
By changing the detection vector $\xi=(\Theta_D,\Delta_D)$, we can affect the number, completeness and reliability of detections. 
}\label{fig:zero}
\end{figure}

We next define the quantities that characterize the detection of signals in observations.
In our method, as an example, 
we use a threshold line on spectra (e.g., a flux or signal-to-noise threshold) and measure the width of features lying above the threshold line.
In this case, the height of the threshold, $\Theta_D$, and the width of features selected by the threshold, $\Delta_D$, 
will be used to define a ``detection'' as shown in Figure~\ref{fig:zero}.
We call the set of parameters used to define the detection the ``detection vector'', denoted by 
$\vec{\xi} = \{\Theta_D, \Delta_D \}$. 

Once source and detection vectors are defined, 
we can run a test for each signal, $\vec{s}$, to determine whether it is ``detected" or ``undetected". 
We denote this test using a ``test function'', $\tau (\vec{s};\vec{\xi})$, which gives boolean results.  
To summarize  
\begin{eqnarray}
\vec{s} & = & \{s_1,s_2,\dotsc\},  \\
\vec{\xi} & = & \{\xi_1,\xi_2,\dotsc\}, \\
\tau(\vec{s};\vec{\xi}) & = & \begin{cases} 1 & \text{if detected}, \nonumber \\ 0 & \text{if not detected}. \nonumber \end{cases} 
\end{eqnarray}
In this mathematical framework, the test function is 
a rule from signal space to boolean values, 0 or 1. 
The test function has a functional dependence on the detection vector, $\vec{\xi}$; 
i.e., different $\vec{\xi}$ can produce different test results. 
This means that by varying $\xi$ we can optimize the detection efficiency, reliability and completeness.

In our analysis, we use a Gaussian profile to represent each source signal. 
Hence, the source vector can be written as $\vec{s}  =  \{s_1,s_2,s_3\}$ 
for $s(\lambda) = s_1 \exp{\big(-\frac{(\lambda - s_2)^2}{2 {s_3}^2}\big)}$, 
where $s(\lambda)$ is a source profile. 

Now we take into account the ``noise'' contribution in our mathematical framework. 
A given detection vector produces one test result on a given single signal. 
To study the effects of noise, we describe the source as an MC ensemble
\begin{equation}
\textrm{MC realizations of } \vec{s} \equiv \{ \sigma_1(\vec{s}), \sigma_2(\vec{s}), \cdots ,\sigma_N(\vec{s}) \},
\end{equation} 
where $\sigma_i(\vec{s})$ is the $i$-th MC realization of $\vec{s}$. 
If we rewrite the test result, $\tau(\sigma_i(\vec{s});\vec{\xi})$, of  $\sigma_i(\vec{s})$ in a simpler form as 
\begin{equation}
\tau_i(\vec{s};\vec{\xi}) \equiv  \tau(\sigma_i(\vec{s});\vec{\xi}),
\end{equation}
we can summarize the presented mathematical representations as  
\begin{itemize}
\item $\vec{\xi}$ : detection vector of parametrized detection criteria,  
\item $\vec{s}$ :  source vector of parametrized source signal,
\item $\sigma_i({\vec{s}})$ : an MC realization (simulated profile) of source vector, $\vec{s}$,
\item $\tau_i(\vec{s};\vec{\xi})$ :  a test result from an MC realization, $\sigma_i({\vec{s}})$.
\end{itemize}
We note that the subscript $i$ represents a single MC instance for a given source vector, $\vec{s}$.

\subsubsection{Completeness}

Now we derive completeness and reliability from the settings above. 
When we produce $N$ MC  realizations for $\vec{s}$, 
we have $N$ test results, $\{\tau_1,\tau_2,\dotsc,\tau_N\}$, for a given $\vec{\xi}$. 
Hence, the detection rate (or completeness) can be defined as  
the ensemble average of these test results  
\begin{equation}
Completeness = D(\vec{s};\vec{\xi})  =  \big<\tau_i(\vec{s};\vec{\xi})\big>_{MC},\label{eq:dpf}
\end{equation}
where the ``MC'' bracket represents the Monte Carlo average. 

To derive reliability, we need to quantify the probability of false detections, i.e., detections resulting from noise.  
If the source signal is a null, $\vec{s} = 0$, all detections are false and the result of contiguous (and perhaps correlated) noise spikes. 
Therefore, the false detection probability is the detection rate of a null signal; i.e.,  
\begin{equation}
F(\vec{\xi})  =  D(\vec{s} = 0;\vec{\xi}). \label{eq:fdf}
\end{equation}
This false detection rate is an intrinsic limitation of any detection criteria.  
A conservative detection vector can reduce this rate but this 
also suppresses the detection of weak but real source signals. 
Detection vectors must be chosen to maximize completeness, while minimizing false detections. 

Reliability is the complement of false detection probability, $1 - F(\vec{\xi})$. 
However, since $1 - F(\vec{\xi})$ is constant, different SNR detections such as 10$\sigma$ or 100$\sigma$ detections 
have the same reliability value, which is not true. 
More correctly, reliability needs to be described 
as a complement of false detection probability density. 

\subsubsection{Reliability}

We assume a source vector, $\vec{s}$, its  
MC ensemble, $\sigma_i(\vec{s})$, 
and its test results, $\tau_i(\vec{s}; \vec{\xi})$.  
According to the test results, we divide each ensemble $\sigma_i(\vec{s})$ into two disjoint ``detected" and ``undetected" 
subsets denoted by $\sigma_i^D(\vec{s})$ and $\sigma_i^U(\vec{s})$.
These two subsets satisfy 
\begin{eqnarray}
\{ \sigma_i(\vec{s}) \} & = & \{\sigma_i^D(\vec{s})\} \cup \{\sigma_i^U(\vec{s})\}, \label{eq:du1} \\ 
\emptyset & = & \{ \sigma_i^D(\vec{s}) \} \cap \{ \sigma_i^U(\vec{s}) \}, \label{eq:du2} \\
D(\vec{s};\vec{\xi})  & = & n\{\sigma_i^D(\vec{s})\} / n\{\sigma_i(\vec{s})\} \label{eq:du3}
\end{eqnarray}
where we use braces to denote the sets collecting their MC instances 
and $n\{ A \}$ represents the number of elements for the set $A$. 
Equation~\ref{eq:du1} -- \ref{eq:du3} are simply a restatement of the definition of the test function; 
the completeness of Equation~\ref{eq:dpf} is now rewritten as Equation~\ref{eq:du3}. 

We use the term, $\sigma_i(\vec{s})$, to represent one simulated MC profile for $\vec{s}$, 
not a parametrized set belonging to signal space. 
We can estimate the corresponding parameters $ \tilde{s}_i $ for each $\sigma_i(\vec{s})$  
by fitting the profile with a Gaussian model. 
We refer to this estimation process as ``reprojection''. 
We need this ``reprojection'' process to represent $\sigma_i(\vec{s})$ using $ \tilde{s}_i$.

To summarize, for a given source vector, $\vec{s}$,  
we have its MC ensemble $\{ \sigma_i(\vec{s}) \}$ of simulated profiles. 
By reprojection, we obtain its reprojected MC ensemble, $\{ \tilde{s}_i \}$, in signal space. 
From this distribution of  $\{ \tilde{s}_i \}$, we can obtain its corresponding probability density function, 
$\tilde{f}(\vec{x}; \vec{s}) $, for $\vec{s}$, where $\vec{x}$ is an independent variable for the density function in signal space. 
All the three processes, MC realization, reprojection, and building probability density function can be depicted as  
\begin{equation}
\vec{s} \rightarrow \{ \sigma_i(\vec{s}) \} \nleftrightarrow \{ \tilde{s}_i  \} \rightarrow \tilde{f}(\vec{x}; \vec{s}). 
\end{equation}
We denote the reprojection process by ``$\nleftrightarrow$'' due to incompleteness of the fitting process. 
Generally, the fitting procedure depends on root-finding algorithm. 
There are inevitable fitting anomalies such as converging to a wrong solution or failing to converge within the limit of root finding trials.  
Therefore, $\{ \sigma_i(\vec{s}) \}$ and $\{ \tilde{s}_i  \}$ are not necessarily mapped in one-to-one correspondence. 
However, when the fitting process shows fair correspondence between $\{ \sigma_i(\vec{s}) \}$ and  $\{ \tilde{s}_i  \}$, 
we can represent $\{ \sigma_i(\vec{s}) \}$ by its density function, $\tilde{f}(\vec{x}; \vec{s})$, as a practical estimate in signal space. 
In this argument,  the detected subset, $\{\sigma_i^D(\vec{s})\}$, can be described by a probability density function in signal space 
by an acceptable fitting performance as 
\begin{equation}
\{\sigma_i^D(\vec{s})\}  \longrightarrow  \tilde{f}_D(\vec{x}; \vec{s}, \vec{\xi}).   
\end{equation}
We call $\tilde{f}_D(\vec{x}; \vec{s}, \vec{\xi})$ the Detection Probability Density Function (DPDF) for $\vec{s}$. 
We note that, by the reprojection process and the definition of DPDF, all observed detections and  quantities derived from MC simulations 
are now plotted in signal space; in our method, the parameter space for a Gaussian profile. 

From Equation~\ref{eq:dpf}, the normalization of DPDF for $\vec{s}$ is $D(\vec{s};\vec{\xi})  = \int \tilde{f}_D(\vec{x}; \vec{s}, \vec{\xi}) d\vec{x}$. 
But, for convenience, we rescale the normalization condition as 
\begin{equation}
1 \equiv \int \tilde{f}_D(\vec{x}; \vec{s}, \vec{\xi}) d\vec{x},   
\end{equation}
where its original normalization is replaced by the conditional probability, $D(\vec{s};\vec{\xi}) \tilde{f}_D(\vec{x}; \vec{s}, \vec{\xi})$. 

The sourceless DPDF, $\tilde{f}_D(\vec{x}; \vec{s}=\vec{0}, \vec{\xi})$, 
is the ``false detection probability density'' and its normalization is the
``false detection probability" in Equation~\ref{eq:fdf},
\begin{eqnarray}
F(\vec{\xi}) & = & \int D(\vec{s} = \vec{0};\vec{\xi}) \tilde{f}_D(\vec{x}; \vec{s} = \vec{0}, \vec{\xi}) d\vec{x}.
\end{eqnarray}
Therefore, the distribution of $\tilde{f}_D(\vec{x}; \vec{s}=\vec{0}, \vec{\xi})$ is the zone of false detections in signal space. 
To derive the final mathematical expression for reliability, we assume an observed spectrum, $\sigma_o$, and its reprojection, $\tilde{o}$. 
As a trivial case of $\tilde{f}_D(\tilde{o}; \vec{s}=\vec{0}, \vec{\xi}) = 0$, any observed detections would be 
deemed fully reliable due to the perfect decoupling of true and false detections. 
Therefore, at least, we can assign reliability = 1 to all observed detections of  $\tilde{f}_D(\tilde{o}; \vec{s}=\vec{0}, \vec{\xi}) = 0$. 

In the case where $\tilde{f}_D(\tilde{o}; \vec{s}=\vec{0}, \vec{\xi}) \neq 0$, we can have a couple of definitions for reliability. 
One of the simplest definitions would be to reject all of those partially reliable detections,  
\begin{eqnarray}
Reliability & = & \begin{cases} 1 & \textrm{if } \tilde{f}_D(\tilde{o}; \vec{s}=\vec{0}, \vec{\xi}) = 0, \nonumber \\ 
0 & \textrm{if } \tilde{f}_D(\tilde{o}; \vec{s}=\vec{0}, \vec{\xi}) \neq 0. \nonumber \end{cases} 
\end{eqnarray}
This definition can be powerful when focusing on detecting strong emission lines. 
However, it is clear that we need a better definition if we hope to detect weaker emission lines as well. 
The quick-and-dirty extension can be to use the profile shape of $\tilde{f}_D(\vec{x}; \vec{s}=\vec{0}, \vec{\xi})$,
\begin{equation}
Reliability = 1 - \tilde{f}_D(\tilde{o}; \vec{s} = 0, \vec{\xi})/f_{max},
\end{equation}
where $f_{max}$ is the maximum value of $\tilde{f}_D(\vec{x}; \vec{s} = 0, \vec{\xi})$. 
In this definition, we can assign fractional reliability to the detections of $\tilde{f}_D(\tilde{o}; \vec{s}=\vec{0}, \vec{\xi}) \neq 0$.  
Since probability density is not probability, this fractional definition is not mathematically correct for reliability.  
To obtain the correct reliability fraction, 
we choose a ``contour'' in the signal space to separate 
``reliable'' from ``unreliable'' detections. 
For instance, we define this contour to be $C(\mu)$ such that:
\begin{equation}
\mu \equiv \int_{C(\mu) \geq \tilde{f}_D(\vec{x}) \geq 0} \tilde{f}_D(\vec{x}; \vec{s} = \vec{0}, \vec{\xi}) d\vec{x}, \label{eq:bisect1}   
\end{equation}
\begin{equation}
1 - \mu = \int_{f_{max} \geq \tilde{f}_D(\vec{x}) \geq C(\mu)} \tilde{f}_D(\vec{x}; \vec{s} = \vec{0}, \vec{\xi}) d\vec{x}, \label{eq:bisect2}  
\end{equation}
where we have the boundary conditions of $C(0) = 0$ and $C(1) = f_{max}$. 
If we choose $\mu=0.99$, then any observed detection falling outside this region has a less 
than 1\% chance of being a false detection; its ``reliability" is therefore $\ge 99$\%.
Therefore, we can define the reliability at $\vec{x}$ as, 
\begin{eqnarray}
\tilde{F}_D(\vec{x}; \vec{s} = 0, \vec{\xi}) & \equiv & \int_{\tilde{f}_D(\vec{x}) \geq \tilde{f}_D(\vec{x'}) \geq 0} \tilde{f}_D(\vec{x'}; \vec{s} = \vec{0}, \vec{\xi}) d\vec{x'}, \label{eq:rel1} \\  
Reliability &=& 1 - \tilde{F}_D(\vec{x}; \vec{s} = 0, \vec{\xi}). \label{eq:rel2}
\end{eqnarray}
In practice, we do not have to calculate $\tilde{F}_D(\vec{x}; \vec{s} = 0, \vec{\xi})$ for all $\vec{x}$ in signal space. 
We only need a reliability threshold for the adopted detection vector and its corresponding contour line such as $C(0.05)$ or $C(0.01)$ for 95\% or 99\% reliability.  

Before we move on to the implementation, we discuss two points related to our mathematical framework. 
The first is about another possible variant of reliability, 
\begin{equation}
Reliability = 1 - F(\vec{\xi})\tilde{F}_D(\vec{x}; \vec{s} = 0, \vec{\xi}). \label{eq:relalt} 
\end{equation}
Generally the false detection probability is very low, such as $F(\vec{\xi}) = 0.01$. 
If the rejection rate of noise signals is very high, 
then the reliability of a detection may be underestimated by the definition in Equation~\ref{eq:rel2}. 
A weighted approach (provided by Equation~\ref{eq:relalt}) may have some advantages in this case. 
However, we choose to ignore any detection that is indistinguishable from noise, 
and hence conservatively adopt the definition in Eqn 18 for the present study.

The second is about a more general interpretation of the DPDF. Since we have focused on the test of detection, 
we have only needed the sourceless DPDF and derived Equation~\ref{eq:rel1}. 
However, we can derive the DPDF, $\tilde{F}_D(\vec{x}; \vec{s} = \vec{s}_1, \vec{\xi})$, for a general source vector, $\vec{s}_1$. 
For an observed detection, $\tilde{o}$, we can calculate $\tilde{F}_D(\vec{x} = \tilde{o}; \vec{s} = \vec{s}_1, \vec{\xi})$ 
as well as $\tilde{F}_D(\vec{x} = \tilde{o}; \vec{s} = 0, \vec{\xi})$. 
As explained before, the latter, $\tilde{F}_D(\vec{x} = \tilde{o}; \vec{s} = 0, \vec{\xi})$, provides the reliability of the detection. 
Similarly, but more interestingly, the former $\tilde{F}_D(\vec{x} = \tilde{o}; \vec{s} = \vec{s}_1, \vec{\xi})$ provides 
the probability that the observed feature, $\tilde{o}$, is originated from the underlying source, $\vec{s}_1$. 
Therefore, the on-source DPDFs provide all the possible candidates of the underlying sources 
for a given observed detection with their own probabilities, though we do not have to use this 
extra information.

By following the basic descriptions above, the automated pipelines we will implement 
can extract detections from spectra and assign the completeness and reliability values to each detection. 
By restricting the reliability or completeness, we can finalize our detections. 
This is the basic frame for our automated detection method. 
The following sections will present the specific steps to build the detection pipelines. 
 
\subsection{Implementation}\label{sec:impl}

\subsubsection{Standardization and Pseudospectrum}

Generally astronomical spectra are imaged on Charge-coupled Device (CCD) cameras. 
One-dimensional spectra are extracted from the 2D images and typically represented by three data arrays of   
wavelength, spectrum profile, and inverse variance with pixel indices. 
Their pixel sampling rate, signal-to-noise ratio (SNR), and flux calibration 
vary according to weather conditions, instruments, and optics. 
Therefore, we need steps to standardize the observation-dependent 
spectra for use with a generic detection algorithm.

In our detection method, we need two requirements of standardization: 
(1) uniform wavelength sampling and (2) normalization of variance to unity. 
The first is necessary to interchange between wavelength scale and pixel scale. 
The pixel scale is an easier and more standard unit to work on data-processing problems, 
while scientific requirements and constraints are written 
in terms of wavelength. 
Our detection method is built on pixel scale. 
The conversion between the two scales is consistent and trivial when the sampling rate is uniform. 
Therefore, for convenience and consistency, uniform wavelength sampling is required in our method.  
In practice, if pixel sampling rates vary within $\pm 10\%$ in a target spectral range, 
the uniform resampling is not necessary. 

The second is to normalize the noise level to unity. 
Since the noise level is a complex result of observational effects 
(e.g., due to instruments, optics, detectors, weather, etc.), we cannot model all possible situations in our MC simulations.
Therefore, we normalize the observed spectra by their standard deviations to equalize the noise to unity. 
That is, if we denote the original data set of wavelength, spectrum, standard deviation of noise 
by $(\lambda(x), s(x), \sigma(x))$, then the normalized data set will be $(\lambda(x), s(x)/\sigma(x), 1)$. 
We call this normalized spectrum, $s(x)/\sigma(x)$, the ``pseudospectrum". 
Our Monte Carlo analysis is built for these pseudospectra. 
The pseudospectrum is not in physical units (e.g., flux density), but is effectively a dimensionless signal-to-noise ratio.
The profile shape in the pseudospectrum is also deformed during the normalization process. 
This deformation is relatively small if the noises in neighboring pixels are similar. 
In practice, the shapes of features in the pseudospectra are similar to those in the original spectra 
as long as the noise varies smoothly under them, as is the case in spectral regions that do not include strong telluric lines.

To describe the meaning of a pseudospectrum, we start by summarizing the quantities in mathematical terms.  
In the spectral range of $(\lambda_0, \lambda_0 + \Lambda)$, 
the flux, pseudoflux (integration of pseudospectrum), and SNR can be written as :
\begin{eqnarray}
\text{Flux} & = & \int_{\lambda_0}^{\lambda_0 + \Lambda} s(\lambda) d\lambda  =  \sum_{x=0}^{N} s(x), \nonumber \\
\text{Pseudoflux} & = & \int_{\lambda_0}^{\lambda_0 + \Lambda} \frac{s(\lambda)}{\sigma(\lambda)}  d\lambda 
= \sum_{x=0}^{N} \frac{s(x)}{\sigma(x)} \nonumber, \\
\text{SNR} &= &\frac{ \int_{\lambda_0}^{\lambda_0 + \Lambda} s(\lambda) d\lambda}
{\sqrt{ \int_{\lambda_0}^{\lambda_0 + \Lambda} \sigma^2(\lambda) d\lambda } }
= \frac {\sum\limits_{x=0}^{N} s(x)}{ \sqrt{\sum\limits_{x=0}^{N} \sigma^{2}(x)}} 
= \frac {\sum\limits_{x=0}^{N} s(x)}{ \sqrt{N} \sigma_{rms}}, \nonumber
\end{eqnarray}
where $\lambda(x=0) = \lambda_0$, $\lambda(x=N) = \lambda_0 + \Lambda$, 
and $\sigma_{rms}$ is a root-mean-square of $\sigma(x)$; generally, $\Lambda$ is a couple of times the FWHM of a typical emission line. 
For the trivial case of uniform noise,  $\sigma(x) = \sigma_0$, 
the mean and rms values are the same, $\sigma(x) = \bar{\sigma} = \sigma_{rms} = \sigma_0$, where $\bar{\sigma}$ is the mean of $\sigma(x)$. 
If we extend this trivial case to ``quasi-uniform'' cases of $\sigma(x) \sim \bar{\sigma} \sim \sigma_{rms}$, 
the pseudoflux can be approximated as  
\begin{equation}
\text{Pseudoflux} \sim \frac{\sum\limits_{x=0}^{N} s(x)}{\bar{\sigma}} \sim \sqrt{N} \times \text{SNR} \label{eq:pseudo}
\end{equation}
Unless our target line emission falls in close proximity to a bright sky line, Equation~\ref{eq:pseudo} generally holds. 
The pseudospectrum is (roughly) a scaled measure of the SNR per pixel. 

\begin{figure*}
\centering
\includegraphics[height=5.5 in]{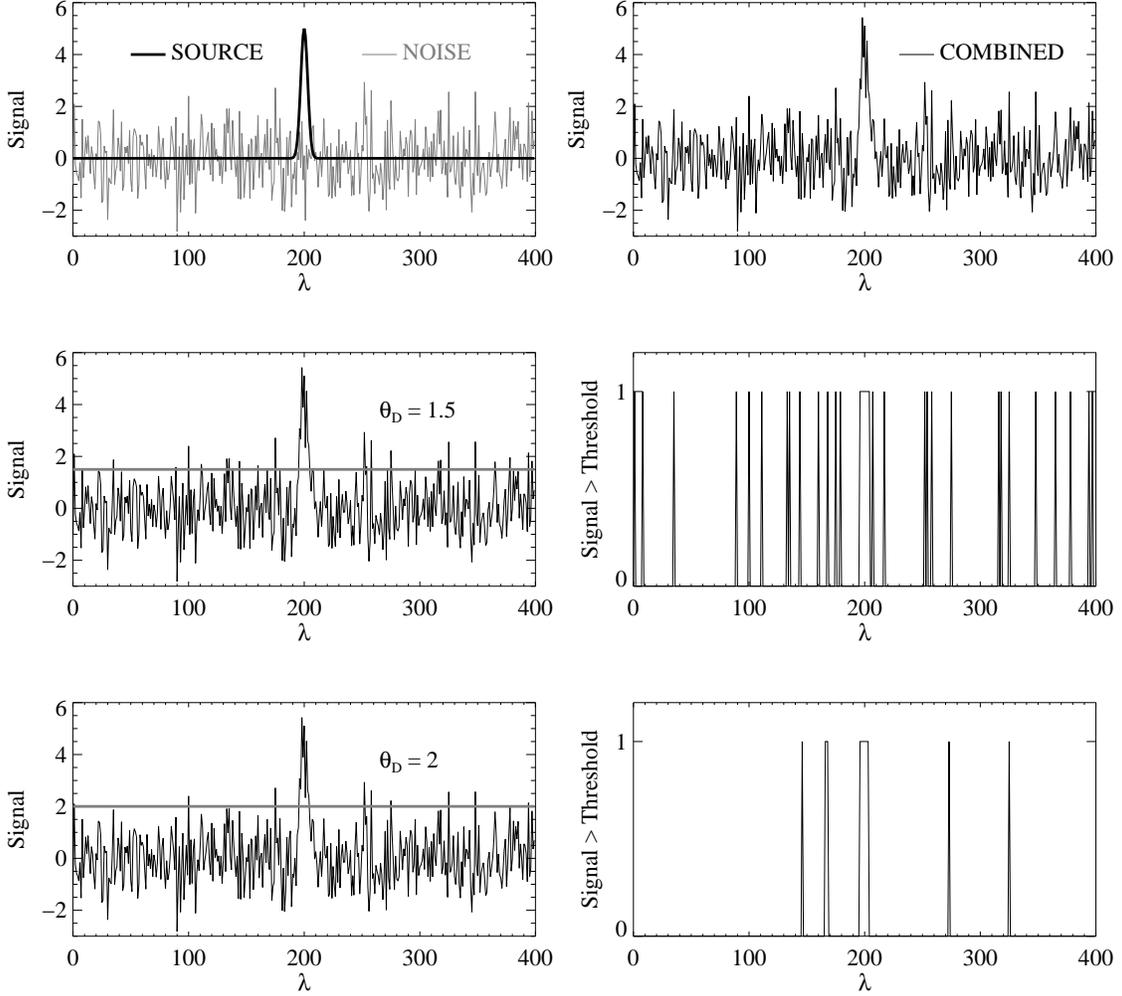}
\caption{The left-top panel shows a Gaussian random noise with $\sigma = 1$ (grey) and a Gaussian source signal 
with $f_{max} = 5$ and $\sigma = 3$ (black). The right-top panel shows the combined profile of the source and noise. 
With the given threshold lines, $\Theta_{D} = $ 1.5 (middle-left) and 2 (bottom-left), 
we can make the detection mask assigning ``1'' where the signal is larger than threshold (middle- and bottom-right panels). 
This detection mask 
results in detection blocks of different widths $\delta$.
For each random realization, we measure detected widths and stack them  
to the probability function, $p_{\lambda} (\delta; \Theta_{D})$, at each pixel position. 
For a given position, $\lambda$, and a detection threshold, $\Theta_{D}$, the sum of all stacked counts for each detected $\delta$ is 
exactly the same as the total number of random realizations; i.e., 
$\sum_{\delta=0}^{\infty} p_{\lambda} (\delta; \Theta_{D}) = 1$. To construct the detected width at $\lambda = 200$ 
is larger than at other sourceless positions. This property is the key feature for our detection method.  
}\label{fig:one}
\end{figure*}

\begin{figure*}
\centering
\includegraphics[height=4.0 in]{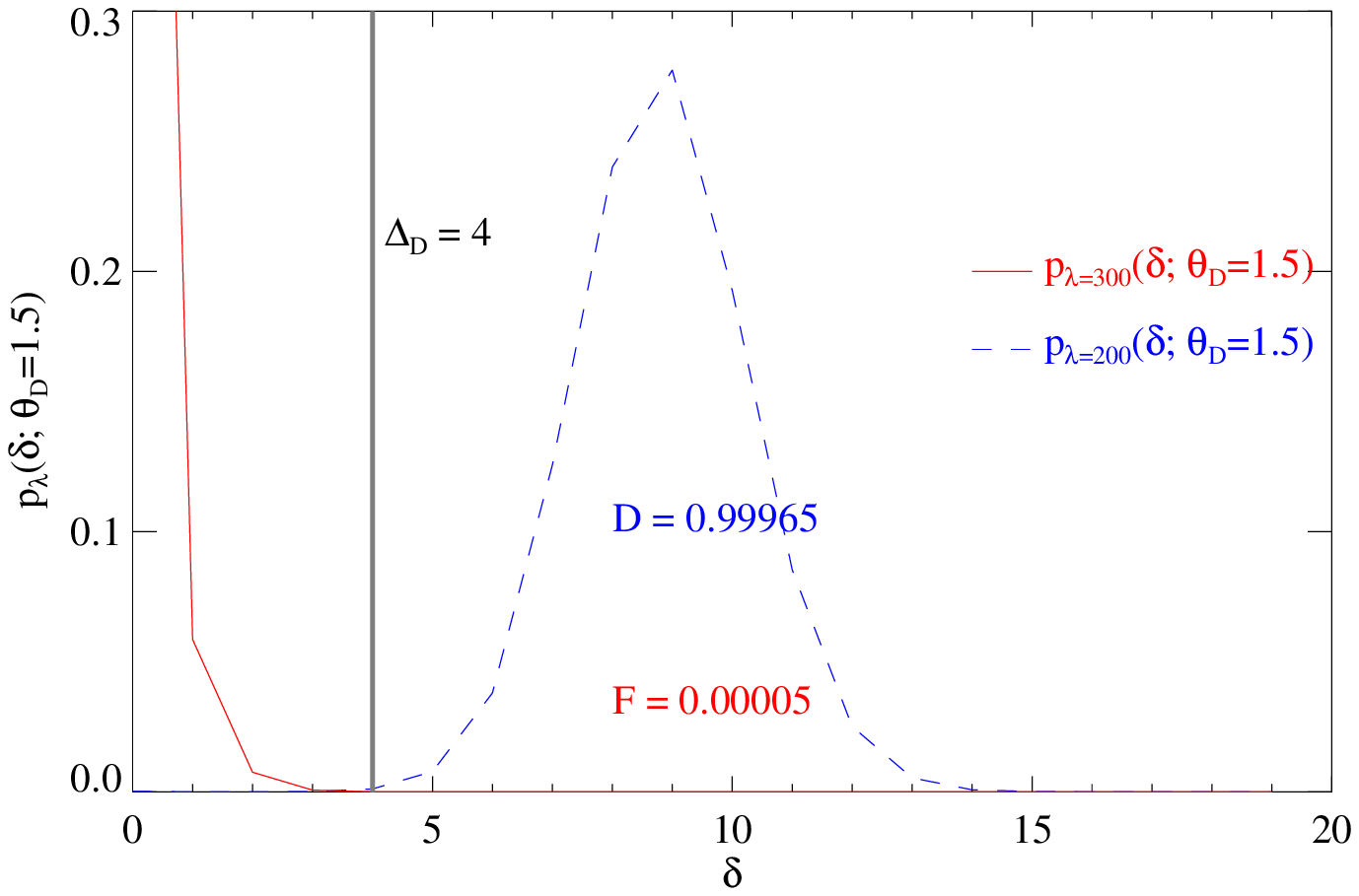}
\caption{The distribution of detected widths, $\delta$, at $\lambda$=200 and $\lambda$= 300 from $10^{5}$ random realizations. 
The measured $\delta$ distribution at the on-source position of $\lambda=200$ is different from 
the one at the off-source position of $\lambda=300$.  
}\label{fig:three}
\end{figure*} 

\subsubsection{Threshold, Detection Mask, and Detected Width}


We now choose a detection threshold and create a boolean 
detection mask from the pseudospectrum: ``1'' when pixel values fall above the threshold; 
``0", when they lie below it.
Figure~\ref{fig:one} shows the two threshold lines, $\Theta_D = $ 1.5 (middle panels) and 2.0 (bottom panels) 
and their detection masks, where $\lambda$ is on a uniformly resampled pixel scale explained in the previous section. 
The top-left panel 
shows a spectrum composed of gaussian random noise with $\sigma=1$ (grey line) 
and a gaussian emission feature of width 3 and height 5 centered at pixel $\lambda=200$ (black line). 
The top-right panel shows their combined signal (pseudospectrum). 
Using the detection mask, we identify some clustered blocks of pixels and measure their sizes (widths). 
We call this block size the ``detected width'', $\delta$. 
The signal produces the largest detection width above the chosen threshold and 
we must choose a detection width to maximize the line detection and avoid false detections caused by the noise spikes.

Hence, for a given source vector $\vec{s}$, we obtain one detected width value from each random noise realization. 
From a series of MC realizations, we obtain an MC ensemble of the detected widths denoted by $\{ \delta_i(\vec{s}, \Theta_D)\}$. 
These measured detected widths are stacked and reduced to the probability function :  
\begin{equation}
\{ \delta_i(\vec{s}, \Theta_D)\} \rightarrow p(\delta; \vec{s}, \Theta_{D}) 
\end{equation}
with the normalization of $\sum\limits_{\delta=0}^{\infty} p(\delta; \vec{s}, \Theta_{D}) = 1$.
We call this probability function as Detected Width Probability Function (DWPF). 

The two ensembles of on-source $\{ \delta_i(\vec{s}, \Theta_D)\}$ 
and off-source $\{ \delta_i(\vec{0}, \Theta_D)\}$ (i.e., on and off the emission line feature) width measurements 
are the fundamental data sets used to define the reliability and completeness in our detection method. 
From the ensembles, we obtain the corresponding DWPFs, $p(\delta; \vec{s}, \Theta_{D})$ and  $p(\delta; \vec{0}, \Theta_{D})$. 
Figure~\ref{fig:three} shows the two DWPFs, $p_{\lambda = 200}(\delta; \Theta_{D})$ and $p_{\lambda = 300}(\delta; \Theta_{D})$, 
measured at the two different positions, $\lambda = 200$ and $300$, for the given source vector $\vec{s}$ shown in Figure~\ref{fig:one}. 
By definition, the DWPF measured at $\lambda = 200$ is the on-source DWPF for $\vec{s}$, 
\begin{equation} 
p(\delta; \vec{s}, \Theta_{D}) = p_{\lambda = 200}(\delta; \Theta_{D}). \label{eq:on}
\end{equation}
We can obtain the off-source DWPF by rerunning the MC simulations for $\vec{s} = \vec{0}$. 
However, since there is no contribution from the source signal at $\lambda = 300$ (or sufficiently away from the signal), 
we can measure the off-source DWPF simultaneously as 
\begin{equation} 
p(\delta; \vec{0}, \Theta_{D}) = p_{\lambda = 300}(\delta; \Theta_{D}). \label{eq:off}
\end{equation}
The clear difference between $p_{\lambda = 200}(\delta; \Theta_{D})$ and $p_{\lambda = 300}(\delta; \Theta_{D})$ shown in Figure~\ref{fig:three} 
is the key feature to set our detection criteria.

\begin{figure*}
\centering
\includegraphics[height=4.0 in]{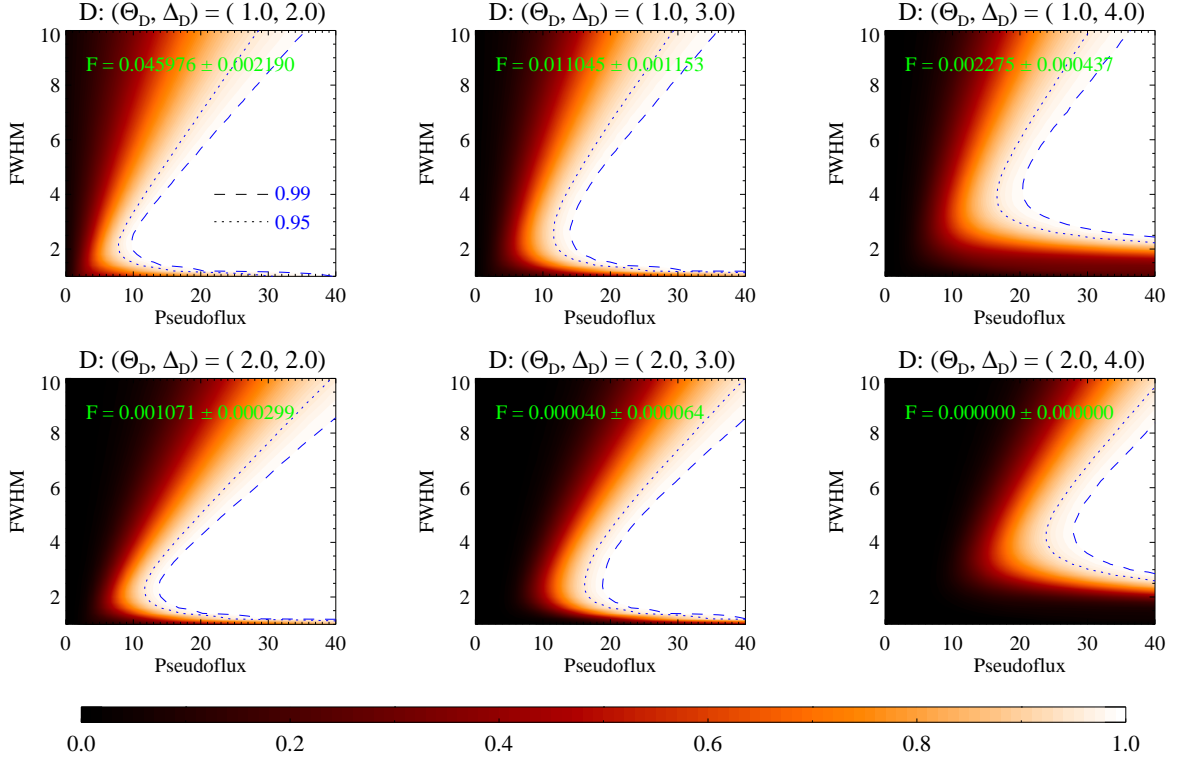}
\caption{The completeness (blue contours) and false detection probabilities (shown in green) from $10^4$ random noise realizations 
for various detection parameters; $(\Theta_D,\Delta_D)$. 
By adopting more conservative detection criteria; i.e., higher $\Theta_D$ or higher $\Delta_D$, we can reduce the false detection probability  
but, at the same time, we lose more completeness; i.e., losing more chances to detect faint source signals. 
We need to consider this trade-off when choosing detection parameters. 
To apply our Monte Carlo results to MMT/Hectospec data, we choose $\Theta_D = 1.5$ and $\Delta_D = 4$. 
}\label{fig:four}
\end{figure*}

\subsection{Completeness} 

From Equation~\ref{eq:on} and \ref{eq:off}, we can directly derive completeness from the definition of ``detection''. 
Figure~\ref{fig:three} shows that the probability of $\delta \ge 4$ for the off-source ensemble is pretty low, 
while most of the on-source ensemble have $\delta \ge 4$. 
From this key trend, we choose a detection width threshold,  $\Delta_D$, 
and define ``detection'' when the measured width is larger than this threshold, $\delta \ge \Delta_D$. 
To summarize, we have the two parameters, $\Theta_D$ and $\Delta_D$, determining our detection criteria. 
If we parametrize a Gaussian profile using its pseudoflux and full-width-half-maximum(FWHM),  
the source vector and detection vector of our method can be denoted by  
\begin{eqnarray}
\vec{s} & = & \{\text{Pseudoflux, FWHM}\},\\
\vec{\xi} & = &\{\Theta_D, \Delta_D\}.
\end{eqnarray}
We now write the detection probability and false detection probability defined in Equation~\ref{eq:dpf} and~\ref{eq:fdf}  as 
\begin{eqnarray}
D(\vec{s}; \Theta_{D},\Delta_D) &=& \sum_{\delta \geq \Delta_{D}} p(\delta;\vec{s}, \Theta_{D}), \\
F(\Theta_{D},\Delta_D) &=& \sum_{\delta \geq \Delta_{D}} p(\delta;\vec{0}, \Theta_{D}).
\end{eqnarray}
using Equation~\ref{eq:on} and~\ref{eq:off}. Figure~\ref{fig:three} shows the detection and false detection probabilities, 
$D(\vec{s}; \Theta_{D} = 1.5 ,\Delta_D = 4) = 0.99965$ and $F(\Theta_{D} = 1.5,\Delta_D = 4) = 0.00005$, 
where the source vector, $\vec{s} = \{38, 7.1\}$, is given on the top-left panel in Figure~\ref{fig:one}. 
Since $D(\vec{s}; \Theta_{D},\Delta_D)$ is the completeness for $\vec{s}$, we finally have 
\begin{equation}
Completeness \equiv \sum_{\delta \geq \Delta_{D}} p(\delta;\vec{s}, \Theta_{D}).
\end{equation}
For each set of detection criteria $\{ \Theta_{D},\Delta_D \}$, we can calculate the completeness values 
for all source vectors $\vec{s} = \{\text{Pseudoflux, FWHM}\}$. 

Figure~\ref{fig:four} shows the completeness contours in the signal space for various $\{ \Theta_{D},\Delta_D \}$, 
derived from $10^4$ MC realizations. 
The blue dotted and dashed lines show 0.95 and 0.99 contours of completeness. 
Strong signals of high pseudofluxes guarantee high completeness as we can expect. 
For the same pseudofllux, the profiles with smaller FWHMs are more likely to be detected, since 
their peaks are higher. The limit of the narrowest width is enforced by the threshold width, $\Delta_D$. 
All of these properties form the bow-shaped contours of completeness in the signal space. 

As $\Theta_D$ and $\Delta_D$ increase, the 0.95 and 0.99 contour lines move to the right side 
reducing the coverage area, but ensuring smaller false detection probabilities. 
This tradeoff needs to be considered when choosing the final detection vector. 
Our data from MMT/Hectospec have the spectral resolution of FWHM$=6$\AA ~and pixel sampling rate of $1.1$\AA/pixel. 
Based on Figure~\ref{fig:four}, we choose $\{\Theta_D,\Delta_D\}=\{1.5,4\}$ as the final detection vector. 
This corresponds to the detection of the core of an emission line 
that has 4 pixels lying at $\ge 1.5\sigma$, or roughly a total core flux of $\ge 3\sigma$.

\begin{figure}
\centering
\includegraphics[height=5.0 in]{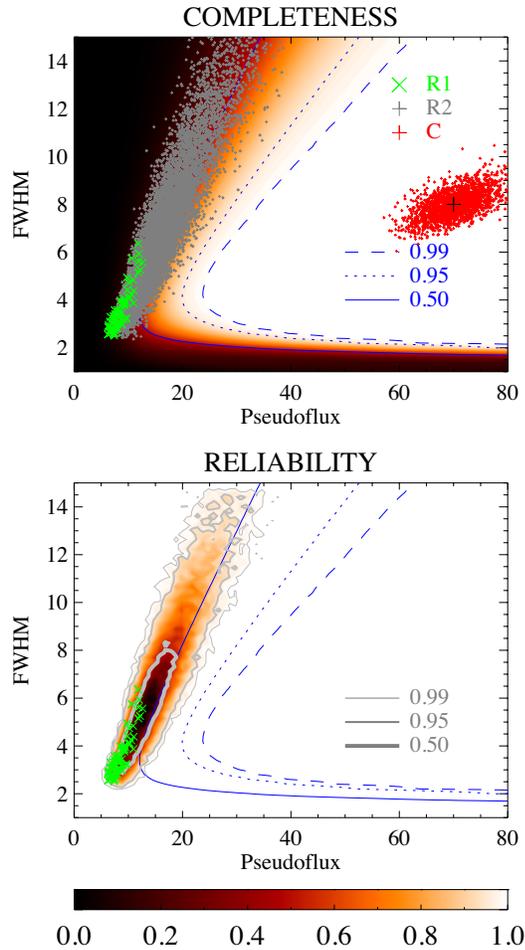}
\caption{Top-panel: The contour plot of completeness for $\Theta_D = 1.5$ and $\Delta_D = 4$ and the reprojections of 
the three samples,  
(1) C: the signal for pseudoflux=70 and FWHM=8, 
(2) R1 : false detections for $\Theta_D = 1.5$ and $\Delta_D = 4$, and  
(3) R2 : false detections resulting from adding + 1 $\sigma$ to the baseline; this results in an effective threshold $\Theta_D = 0.5$. 
R2 is an upper bound for the false detections; 
in practice, the false detection distribution is likely to lie between the R1 and R2 distributions.
Bottom-panel: The grey contours show the R2 distribution. The region lying outside a given R2 contour (in grey) is 
reliable at the contour levels on the bottom panel.
}\label{fig:five}
\end{figure}

\subsection{Reliability} \label{sec:r}

As described in \S \ref{sec:bi}, we need to investigate each MC realization to derive reliability. 
Since we have one detection width for one MC realization, we can use this one-to-one correspondence  
to represent each MC realization, 
 \begin{equation}
\vec{s} \rightarrow \{ \sigma_i(\vec{s}) \} \thicksim \{ \delta_i(\vec{s}, \Theta_D)\} 
\end{equation}
Then, the detected and undetected subsets also can be represented by the two groups of $\delta \ge \Delta_D$ and $\delta < \Delta_D$ as 
\begin{eqnarray}
\{ \sigma_i^D(\vec{s}) \} \thicksim \{ \delta_{\ge \Delta_D}(\vec{s}, \Theta_D)\}, \\
\{ \sigma_i^U(\vec{s}) \} \thicksim \{ \delta_{< \Delta_D}(\vec{s}, \Theta_D)\}.
\end{eqnarray}
Finally, the reprojection is the Gaussian fit on the feature where $\delta \ge \Delta_D$, 
\begin{equation}
\{ \delta_{\ge \Delta_D}(\vec{s}, \Theta_D)\} \longrightarrow \{ \tilde{s}_{\ge \Delta_D} \}. 
\end{equation}
The top panel of Figure~\ref{fig:five} shows three reprojection samples for 
\begin{itemize}
\item C: the signal of pseudoflux=70 and FWHM=8, 
\item R1 : false detections for $\Theta_D = 1.5$ and $\Delta_D = 4$, 
\item R2 : false detections adding + 1 $\sigma$ to baseline; its effective threshold $\Theta_D = 0.5$, 
\end{itemize}
with the completeness contours for $\Theta_D = 1.5$ and $\Delta_D = 4$. 

The completeness of the sample C (the red points) is equal to 1. 
The most probable point (black cross) of C coincides with its original source signal of pseudoflux=70 and FWHM=8. 
This is a typical reprojection pattern for strong signal. 
The R1 sample shows the reprojections of false detections, i.e., zero source signal, for $\Theta_D = 1.5$ and $\Delta_D = 4$. 
This is the theoretical limit of false detections. 
Practically, however, 
baseline estimation is rarely perfect and we may need to allow for some additional systematic uncertainty. 
Typically, under-/overestimating the baseline results in under-/overestimating the false detection rate.
Since the suppressed case shows lower false detection rate than the theoretical limit, 
the more problematic case, though both are problematic,   
is when the baseline is overestimated to enhance false detections producing a larger unreliable region on the signal space.  
We need to take into account these overestimated false detections 
caused by baseline uncertainty.  
As an upper limit of the false detections,  
we add  $+1\sigma$ to the baseline and measure the detected widths; its effective detection threshold is $\Theta_D = 0.5$. 
This is the R2 sample. This generous upper limit may cover the conditions of poor sky subtraction or continuum subtraction 
which increase the uncertainty of baseline estimates. 

The bottom panel of Figure~\ref{fig:five} shows the final reliability contours defined in Equation \ref{eq:rel1} and \ref{eq:rel2} 
for the R2 sample. This panel is our key result, and is used to assign reliability and completeness to each detected feature. 
Since R2 is an upper limit, the R2 contours provide more conservative reliability (i.e., more generous false detections probability density). 
The actual distribution of false detections likely lie between R1 and R2.

\begin{figure*}
\centering
\includegraphics[height=2.0 in]{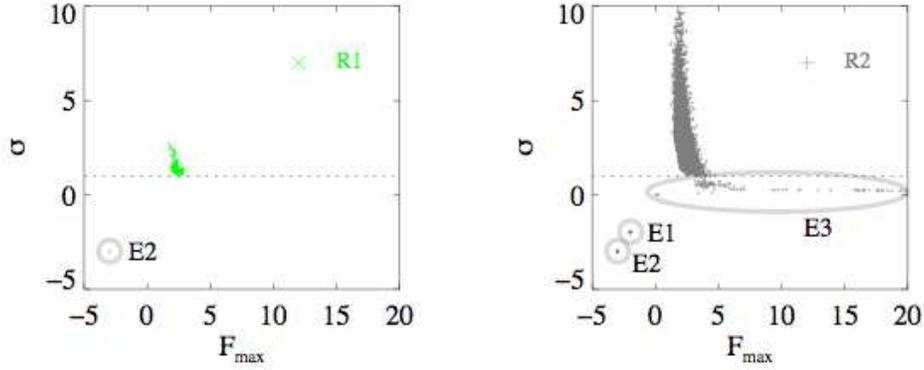}
\caption{The fitting results for R1 and R2 samples, demonstrating the incomplete reprojection process. 
E1 and E2 are artificially assigned to $(-2,-2)$ and $(-3,-3)$. 
E3 has the width smaller than 1 pixel size, which is under Nyquist's sampling. 
}\label{fig:six}
\end{figure*} 

\subsection{Incomplete Reprojections : Fitting Anomalies}
As previously mentioned, most root finding or optimization routines suffer from the issues 
of converging to local minimum (finding a wrong solution) or never converging to any solution (failing to find solution). 
The reprojection process, therefore, is not complete due to these inevitable fitting anomalies.
Here, we investigate these anomalies to quantify the reprojection process. 

In our method, we use MPFIT packages (Markwardt 2009) 
to fit a Gaussian line profile to each detected feature in order to estimate the source parameters. 
For a given detected width, $\delta$, 
we add a width, W, on both of left and right sides 
of the 
detection to define a fitting range of $\delta+2W$.
When we run MPFIT with default settings with W = 3, 
we characterize three kinds of fit anomalies : (1) E1: centroid is out of the fitting range, (2) E2: negative flux, and (3) E3: 
width is narrower than 1 pixel length (which is not possible for real features given the oversampled data).
Figure~\ref{fig:six} shows these fitting results for the R1 and R2 samples. 
Instead of pseudoflux and FWHM, we use the height, $F_{max}$, and Gaussian sigma width, $\sigma$, to show the anomalies more clearly. 
They are related as pseudoflux = $2.5066 \sigma F_{max}$ and FWHM = $2.3548\sigma$. 
Because we search the emission lines within the detected width, all of the three anomalies are not valid fits. 

In order to minimize the fitting anomalies, we investigated the impact of various constraints 
to the post-detection line-fitting algorithm and converged 
on the following constraints: (1) positive flux only, (2) W = 3, 
and (3) the centroid must be located in the fit range. 
Based on these constraints, we obtained near-complete reprojections; 
zero anomalies for the R1 sample and four E3 anomalies 
for the R2 sample from $10^6$ MC realizations. 

\begin{figure*}
\centering
\includegraphics[height=3.0 in]{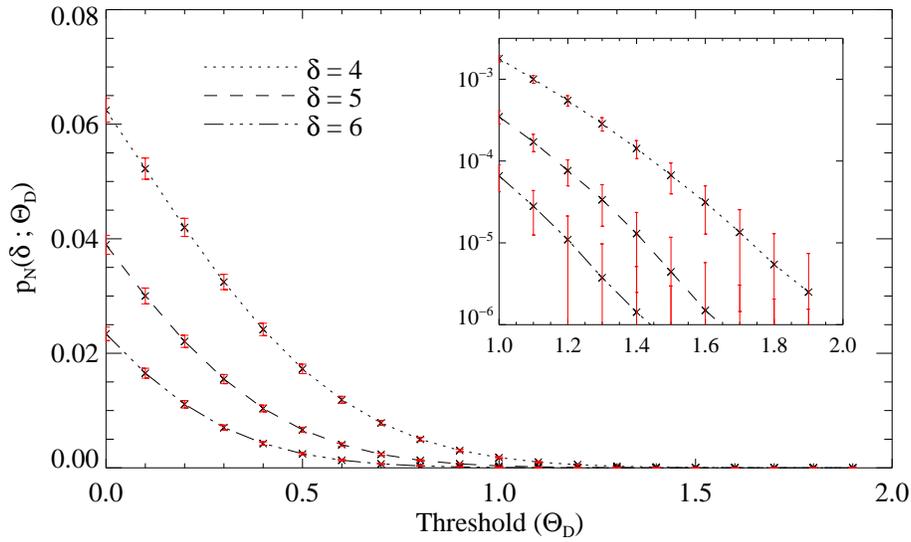}
\caption{The noise DWPF derived from $10^5$ MC realizations. 
The inset figure shows the noise DWPF in log scale for $\Theta_D \ge 1.0$. 
For $\Theta_D=1.5$ and 0.5 and a minimum detection width of $\delta=4$, 
the binomial distribution sums (Equation~\ref{eq:numstat}) suggest 
false detection rates of $\approx$0.0002\% and 0.6\% respectively.
}\label{fig:seven}
\end{figure*} 

\begin{figure}
\centering
\includegraphics[height=4.0 in]{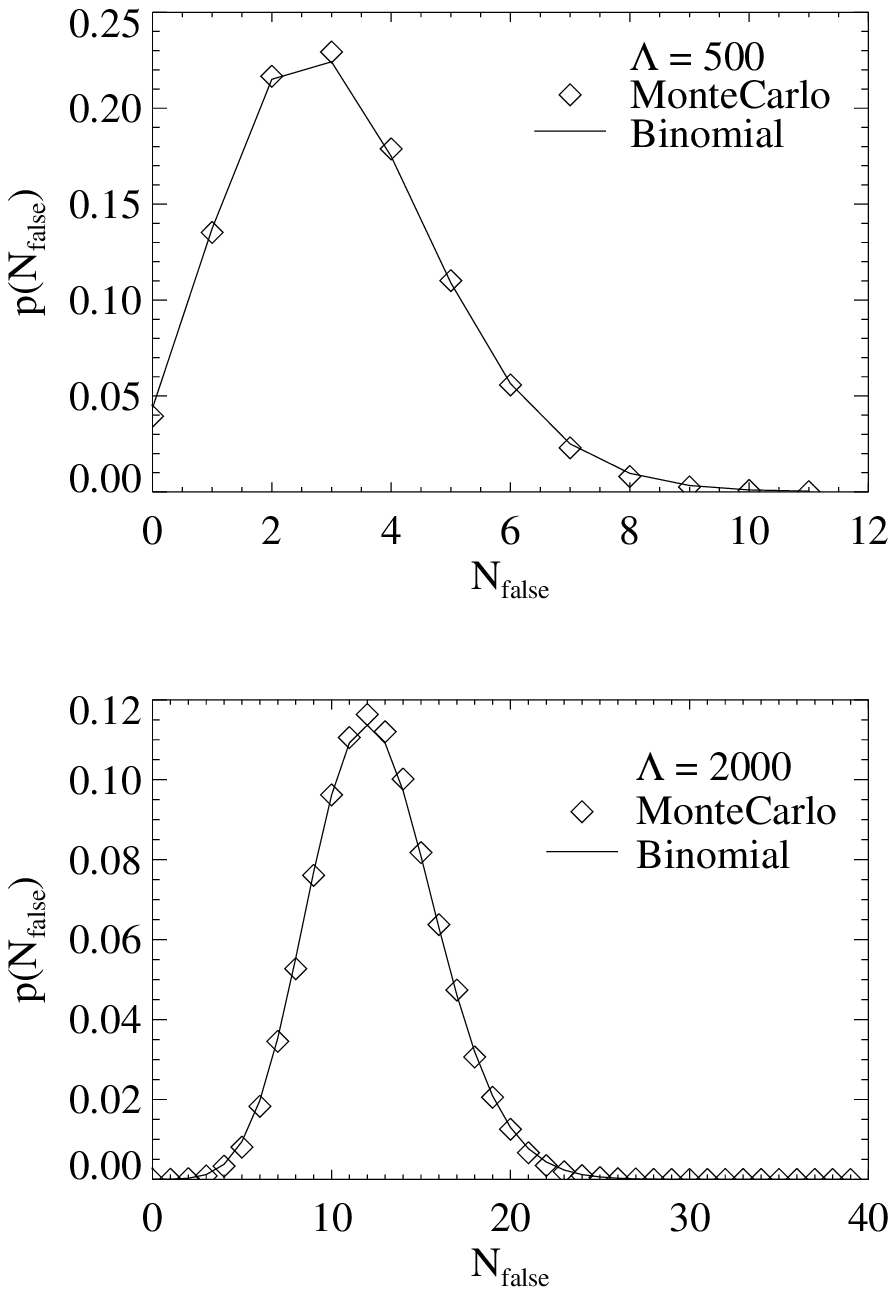}
\caption{ The number statistics of false detections for the two spectral ranges; 
$\Lambda =$ 500 (top) and 2000 (bottom), with the detection criteria $\Theta_D = 0.5$ and $\Delta_D = 4$.  
The diamonds represent the false detection counts from $10^5$ MC realizations, 
and the solid lines the binomial distributions from probabilities derived from DWFP. 
This figure verifies that the number statistics of false detections follows the binomial distribution 
derivable from DWPF; $N_{false} \sim B (\Lambda, \sum\limits_{\delta \ge 4} \frac{p_N(\delta)}{\delta})$. 
}\label{fig:eight}
\end{figure}

\subsection{Number statistics of false detections}\label{sec:numstat}

In this section we discuss how the false detection statistics depend upon the size of the spectral range 
over which the detection search algorithm is run.
We assume that $\Lambda$ is a spectral range of interest and $p_N(\delta; \Theta_D)$ is a noise DWPF as described by Equation~\ref{eq:off}. 
Since the source signal term is zero, $p_N(\delta; \Theta_D)$ only depends on $\Theta_D$. 
Figure~\ref{fig:seven} shows $p_N(\delta; \Theta_D)$ vs. $\Theta_D$. 
Basically, $p_N(\delta; \Theta_D)$ is a probability showing the occurrence of noise detections 
of a width $\delta$ at a certain sourceless position. Therefore, through the range of $\Lambda$, we can conjecture that 
the possible false detection counts are the statistic of binomial trials using the probability of $p_N(\delta; \Theta_D)$. 
This is true if we take into account the redundant counts of $\delta$  
 by diving $p_N(\delta)$ by $\delta$. 
Therefore, the number of false detections in the given spectral range, $\Lambda$, 
can be expressed as : 
\begin{displaymath}
{N}_{false} (\delta; \Theta_{D}, \Lambda) \approx B(\Lambda, \frac{p_N(\delta; \Theta_D)}{\delta})
\end{displaymath}
\begin{equation}
{N}_{false} (\Delta_D, \Theta_{D}, \Lambda) \approx B(\Lambda, \sum\limits_{\delta \ge \Delta_D} \frac{p_N(\delta; \Theta_D)}{\delta}) \label{eq:numstat}
\end{equation}
where $B(N,p)$ represents a binomial distribution of $N$ trials with a probability $p$.

To verify our argument, we produced $10^5$ MC  realizations with $\Lambda = $ 500 and 2000 and measured 
the number counts of noise detections and compared these with their binomial distributions. 
Figure~\ref{fig:eight} shows the false detection counts from MC realizations (diamonds) and the binomial distributions (shown by the lines) 
from Equation~\ref{eq:numstat} for $\Theta_D=0.5$ and $\Delta_D = 4$. We use $\Theta_D=0.5$ to obtain enough counts for comparison, 
since $p_N(\delta;\Theta_D=1.5)$ is too small. 
This good agreement is not a surprising result because $p_N(\delta;\Theta_D)$ itself is derived from the same setups of the MC simulation. 
This binomial description for the number statistics of false detections is in a sense rephrasing the definition of DWPFs. 
If $\Lambda \approx \delta$, the binomial description fails because within the length of $\delta$ the probability is not independent. 
But, in general, the searching spectral range is much larger than the width of emission line, $\Lambda >> \delta$.  
Therefore, for most cases, without running MC simulation for the searching spectral range, 
we can easily estimate the false detection counts from DWPF. 

For example, when we search LAEs 
in the range from 4000 to 5000\AA ~ with 1\AA/pixel sampling, i.e., $\Lambda = 1000$, 
with the detection criteria, $\Theta_D = 1.5$ and $\Delta_D = 4$, the false detection probability, 
$\sum\limits_{\delta \ge \Delta_D} \frac{p_N(\delta; \Theta_D)}{\delta})$, is $1.78\times10^{-5}$. 
Hence, the average false detection counts for each spectrum are $0.0178 \pm 0.1334$. 
Because the false detection counts are much lower than 1 for a single spectrum, 
the detection threshold is good enough to suppress most of the false detections. 
Since our total number of sample spectra is near 1000, a dozen false detections can occur in our sample. 
However, since those false detections have low reliabilities, we can filter them out. 
For  $\Theta_D = 0.5$ and $\Delta_D = 4$, $\sum\limits_{\delta \ge \Delta_D} \frac{p_N(\delta; \Theta_D)}{\delta}$, is 0.00624. 
In this case, we can expect $6.24 \pm 2.49$ false detections for each single spectrum. 
As described in the previous sections about the R1 and R2 samples, these expected counts can be used as  
a generous upper limit for very poor sky subtractions or continuum subtractions.

\section{Application}

We now apply our method to search for Ly$\alpha$ 
emission lines in our our MMT/Hectospec data. 
We have identified LAE candidates from the survey 
using Subaru/SupremeCam with the IA445 filter (Prescott et al. 2008). 
As a spectroscopic followup, we have obtained 1574 spectra from MMT/Hectospec observations (Dey et al. in prep). 
We apply our detection pipelines to those 1574 spectra and describe the results below.
  
\subsection{Data}

\subsubsection{Sample selection}

We performed an intermediate-band survey using 
Subaru/SupremeCam with IA445 filter (Prescott et al. 2008) 
on the NOAO Deep Wide-Field Survey Bo\"otes field (NDWFS; Jannuzi \& Dey 1999). 
The central wavelength of IA445 is 4458 \AA~and its FWHM width 201 \AA. 
The full details about the reductions of the IA445 image are presented in Prescott et al. and Dey et al.
Briefly, the total exposure time was 3 hours. The images were reduced using the SDFRED software 
(Yagi et al. 2002; Ouchi et al. 2004). 
We retrieved $\approx 38600$ sources 
from the IA445 image using SExtractor with 
the $5\sigma$ limit of 26.5 AB mag (Bertin \& Arnouts 1996). 
We registered and resampled IA445 to match the $B_W$ and  $R$ imaging from NDWFS. 


\subsubsection{MMT/Hectospec observations}

We observed the selected candidates using MMT/Hectospec to confirm and obtain spectroscopic redshifts of LAEs.  
Hectospec is a 300-fiber and $1^\circ$ field-of-view multiobject spectrograph at the 
MMT Observatory (Fabricant et al. 2005). We used the program XFITFIBS to assign optical fibers 
to science and calibration targets in our field and observed 7 configurations. 
We used the 600 line/mm grating blazed at 6000\AA with a resolution of FWHM$\approx$6.2\AA. 
We used the 270 line/mm grating blazed at 5200\AA\ with a resolution of FWHM $\approx$ 6.2\AA.

We observed 1574 LAE candidates using 7 configurations over 6 nights as summarized 
in Table 1. The seeing was $> 1.6$\arcsec in windy weather on 5/16 and 6/17 in 2012. 
The effective exposure times for M2 and J1 are, therefore, smaller than the presented ones 
and their data qualities are poor. 
On the other days, the weather conditions were acceptable and the seeing $0.64 - 1.0$\asec. 
The details about the MMT/Hectospec observations and reductions will be presented by Hong et al. (2013; in prep). 
Briefly, we use the HSRED package, a modified version of the Sloan Digital Sky Survey (SDSS) pipeline, 
written in Interactive Data Language (IDL) by Richard Cool\footnote{http://www.mmto.org/\~{}rcool/hsred/}. 
A description of the HSRED also can be found in Kochanek et al. (2012). 
The basic reduction of subtracting bias, flat-fielding, calibrating arc, and sky subtraction are done by the HSRED. 
We apply our detection method to the output spectra from the HSRED pipelines.

\begin{deluxetable*}{crrrrr}
\tabletypesize{\footnotesize}
\tablecolumns{6}
\tablewidth{0pc}
\tablecaption{MMT/Hectospec observations}
\tablehead{
\colhead{Configuration}& \colhead{R.A.\tablenotemark{a}} & \colhead{Decl.\tablenotemark{a}} & \colhead{Date\tablenotemark{b}}& \colhead{Exposure Time} 
&\colhead{\# of LAE targets\tablenotemark{c}}  \\
\colhead{(label)} &\colhead{(J2000.0)} &\colhead{(J2000.0)} & \colhead{} & 
\colhead{(sec)} &\colhead{}   
}
\startdata
PM1  & 14:32:23.6 &+33:24:44 & 5/28/2009 & 6$\times$1800& 244 \\
PM2   & 14:33:35.9 &+33:27:16 & 5/29/2009  &  6$\times$1800& 238 \\
\cline{1-6}\\
M1   & 14:32:19.7 &+33:24:05 & 5/17/2012 &  6$\times$1560& 226 \\
M2   & 14:32:59.8 &+33:23:13 & 5/16/2012 &  6$\times$1600&  221\\
   & & & 5/17/2012 &  2$\times$1600&   \\
M3   &14:32:43.1&+33:22:25 & 5/21/2012 & 3$\times$1800 & 232 \\
J1  & 14:32:24.9	&+33:25:31 & 6/17/2012 & 4$\times$1800& 217  \\
  & & & 6/20/2012 & 2$\times$1800&    \\
J2  & 14:33:36.2	&+33:23:24 & 6/21/2012 & 3$\times$1800& 203 \\
\enddata
\tablenotetext{a}{ The pointing position for each configuration. The field of view is $1^{\circ}$ in diameter centered on this position. } 
\tablenotetext{b}{ The seeing was $> 1.6$\asec on windy weather on 5/16 and 6/17 in 2012. The effective exposure times 
for M2 and J1 are, therefore, smaller than the presented ones.  On the other days, the weather conditions were acceptable and the seeing $0.64 - 1.0$\asec. } 
\tablenotetext{c}{ The number of fibers assigned to the LAE candidates out of the total 300 fibers. The total is 1581 with 7 duplicated targets. } 
\end{deluxetable*}

\subsection{Results}

We split the automated processes into two phases; (1) raw detections and (2) customized selections from the raw detections. 
During the first phase, the pipelines catalog all detected features with their own reliability and completeness values. 
In the second phase, we select a reliability threshold to identify ``reliable" detections and maximize exclusion of false positives.

\subsubsection{Raw detections}

For illustration purposes, we start by focusing on a single pointing - the PM1 configuration.
We observe 244 LAE candidates and find 179 spectra having raw detections 
(automated dumps) in the search wavelength range, 4000 -- 5000 \AA, for the criteria of $\Theta_D = 1.5$ and $\Delta_D = 4$. 
The number statistics of false detections for R1 (in \S \ref{sec:r}), 
representing the case of perfect baseline estimation, is $0.016 \pm 0.13$; for all 244 spectra, $3.9 \pm 2.0$. 
For R2 representing the case of poor baseline estimation, 
we expect $5.6 \pm 2.4$ false detections for each spectrum and $1370 \pm 37$ for all 244 spectra. 
Roughly, therefore, we can expect 0 -- 7 false detections on each spectrum. 
However, since the sample R2 assumes that the baseline is overestimated by +1$\sigma$ through the whole searching spectral range, 
the estimate from R2 is an exaggerated upper-bound.  
When the real baseline errors are suppressed or enhanced through the spectral range larger than $\pm1\sigma$, 
the effective spectral length enhancing the +1$\sigma$ baseline error should be some fraction of the total spectral range.  
Therefore, accounting for this effective fraction of the enhanced fluctuation, 
the practical estimates of false detections may be around 0 -- 3 false detections on each spectrum. 

\begin{figure*}
\centering
\includegraphics[height=3.5 in]{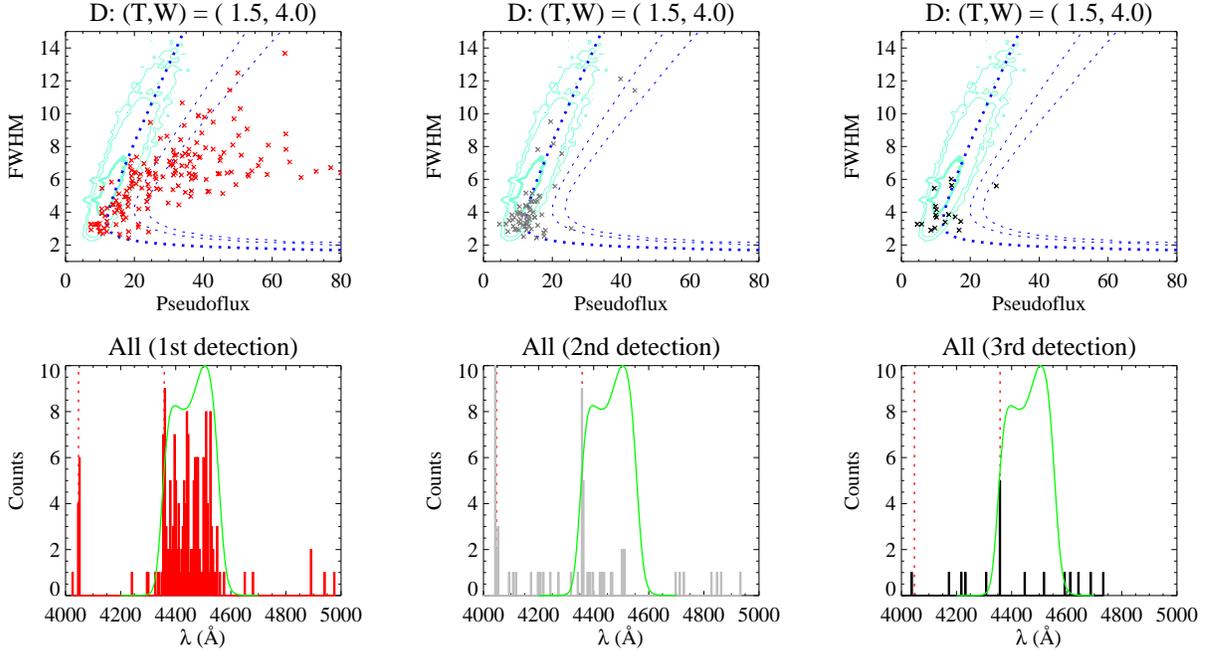}
\caption{ The raw detections resulting from our analysis of the data from MMT/Hectospec configuration PM1.
The top panels show the observed source vectors 
(Gaussian fits of the detections) for the first, second, and third detections in each spectrum, as defined in the text, 
with the reliability and completeness contours shown in Figure~\ref{fig:five}. 
The bottom panels show the histograms of the centroids of the corresponding top panels  
with the green lines representing the throughputs of the IA445 filter. 
The red dotted lines are the sky emission lines, \iion{Hg}{i}  4047\AA~ and 4358\AA. 
}\label{fig:nine}
\end{figure*}

The top panels in Figure \ref{fig:nine} 
show the reprojections (Gaussian fits) of the raw detections  
with the reliability (cyan) and completeness (blue) contours shown in Figure~\ref{fig:five}. 
These top panels are the key outputs from our detection method. 
The bottom panels show the histograms of line centroids with the transmission curve of the IA445 filter (green lines). 
From the false detection statistics presented above, we generally expect a couple of detections on each spectrum. 
When there are more than one detection on each spectrum, 
we call them ``second detections'' for secondary features and ``third detections'' for tertiary features.  
The red points and histogram represent the first detections (the first column), 
the grey the second detections (the second column), and the black the third detections (the third column). 
The red dotted lines on the bottom panels represent the sky emission lines, \iion{Hg}{i}  4047\AA~ and 4358\AA. 

We find two important results from Figure ~\ref{fig:nine}.  
First, the centroid histogram of the first detections (bottom-left panel) shows 
a good correlation with the IA445 filter transmission curve, 
while the other histograms of the second and third detections (bottom-middle and -right panels) do not.
This good match for the first detections implies that 
the target selection and our automated method work properly, suggesting that the first detections are likely to be real emission line detections. 
Conversely, the poor match for the second and third detections implies that they are more likely to be noise detections. 
The fact that their reprojections shown in the top-middle and -right panels fall within the low reliability contours 
also supports this argument.  
A slight excess in the second detections near 4500\AA~   
implies that some of the second detections could be real Ly$\alpha$ emission lines. 
There are hardly any real detections among the third detections. 
Hence, we exclude all the third detections from the real emission candidates. 
Second, there are many junk detections caused 
by residuals associated with the improper subtraction of the \iion{Hg}{i}  4047\AA~ and 4358\AA\ telluric emission lines
in all of the first, second, and third detections. 
The second phase needs to deal with these junk (sky residual) detections. 

\begin{figure*}
\centering
\includegraphics[height=5.0 in]{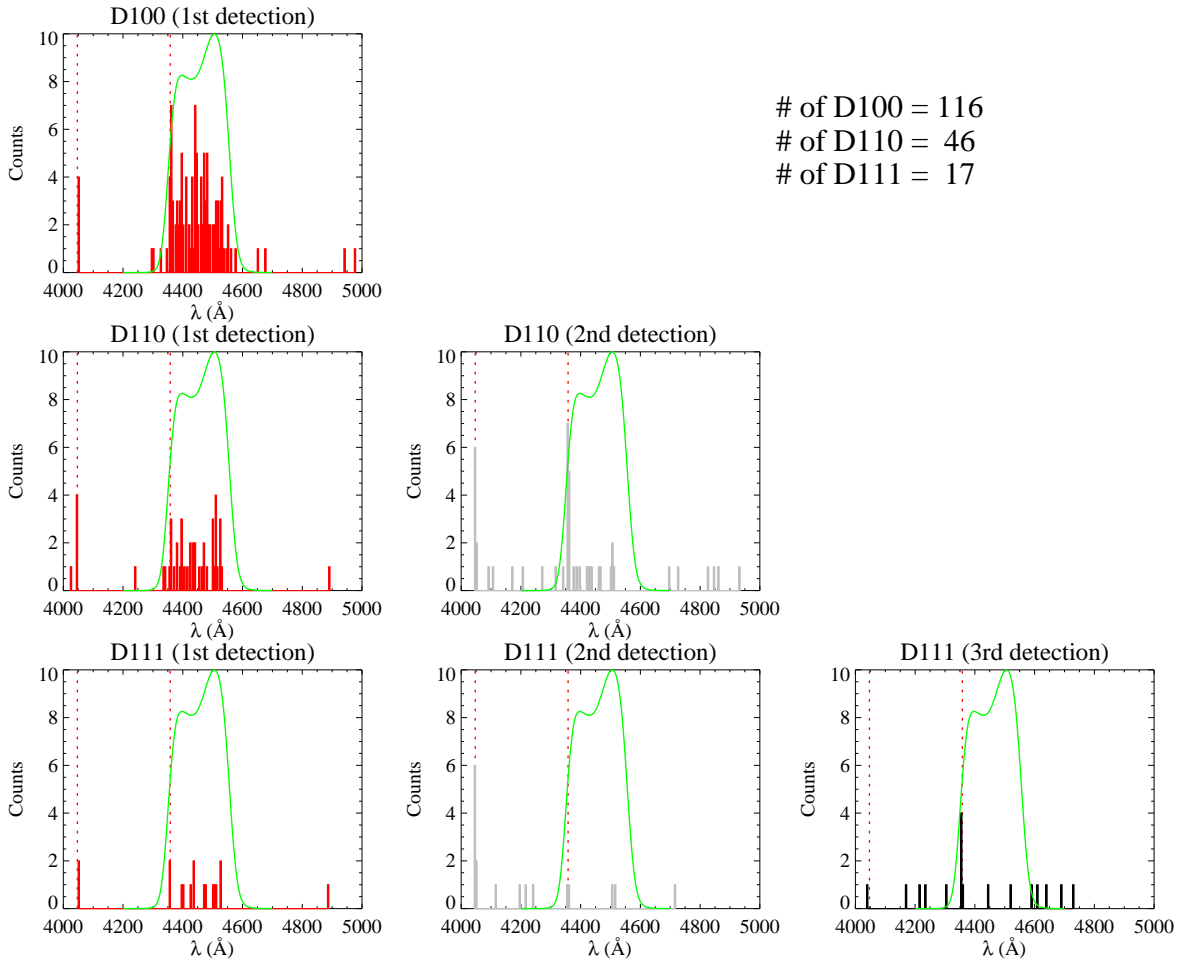}
\caption{ The histograms of the detected centroids for the three detection categories, D100, D110, and D111, in PM1. 
The sum of each column results in the histograms in Figure~\ref{fig:nine}. 
Because the number of false detections is $3.9 \pm 2.0$ for perfect sky and continuum subtractions, 
46 and 17 detections of D110 and D111 are mostly due to erroneous baseline fluctuations 
caused by the practical limitations of sky subtraction.
}\label{fig:ten}
\end{figure*}

To investigate the raw detections further, we categorize the 179 raw detections into three kinds, called D100, D110, and D111.  
D100 represents the spectra having only first detections,  
D110 having first and second detections, and D111 having more than three detections. 
Figure~\ref{fig:ten} shows the histograms of the three categories, D100, D110, and D111. 
The sum of each column results in the histograms of Figure~\ref{fig:nine}. 
The numbers of D110 and D111 detections are 46 and 17. 
If we assume there is one real emission line in D110 and D111, 
we have $46+2\times17$ false detections. 
When comparing these 80 false positives with the numbers predicted 
from the R1 and R2 reliability contours ($4\pm2$ and $1370\pm37$ respectively), 
we find that the R2 contour significantly overestimates the false detection rate. 



\begin{figure*}
\centering
\includegraphics[height=7.5 in]{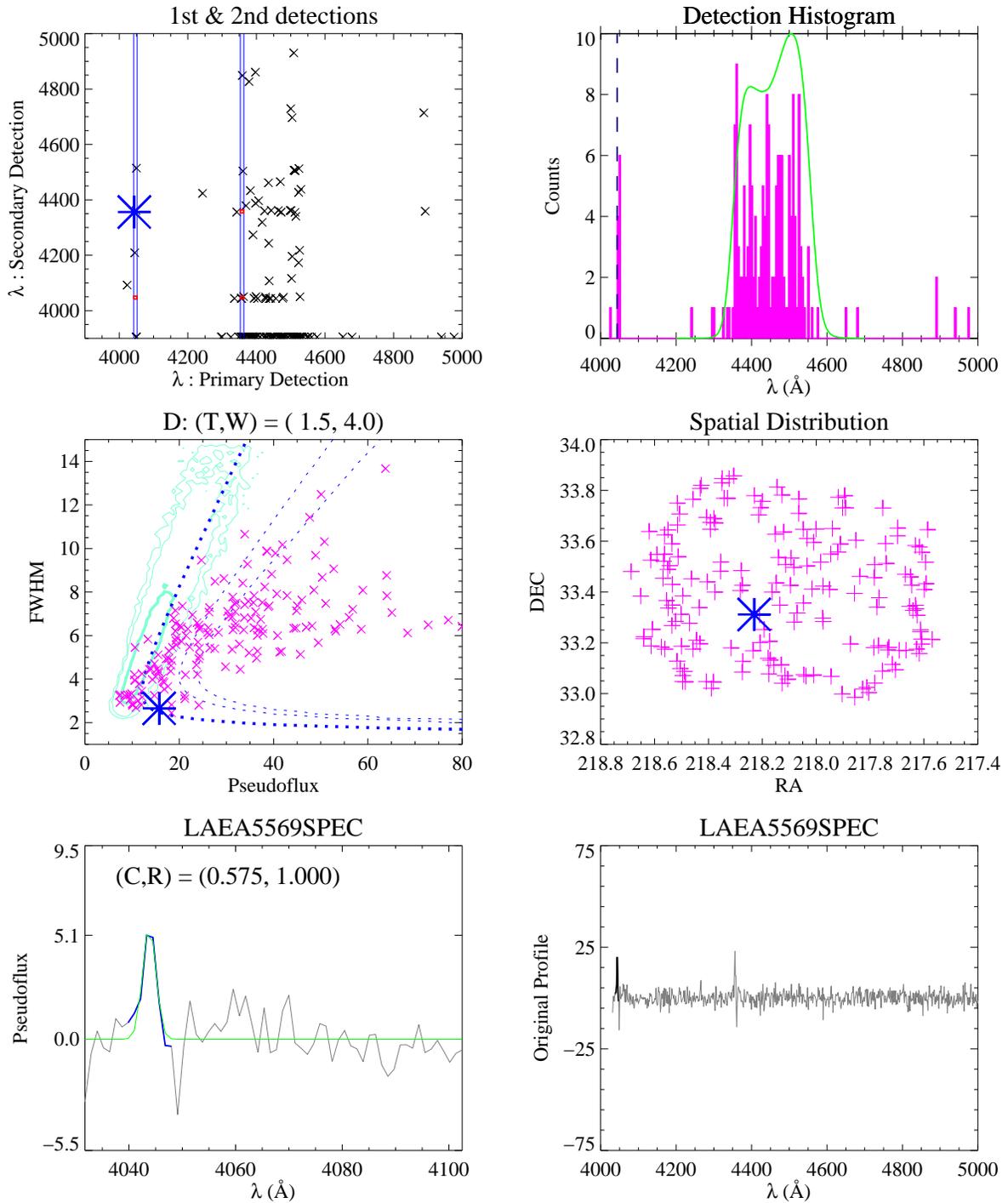}
\caption{An automated raw detection for the target LAEA5569 in the PM1 configuration. 
The blue asterisks and blue dashed vertical line in the top and middle panels show where LAEA5569 is located and  
the two bottom panels show the pseudoprofile (left) and original profile (right) for LAEA5569. 
For this candidate, the first and second detections are sky residuals. This is a typical example needed to be removed 
by appropriate rejection. 
}\label{fig:eleven}
\end{figure*} 

\begin{figure*}
\centering
\includegraphics[height=7.5 in]{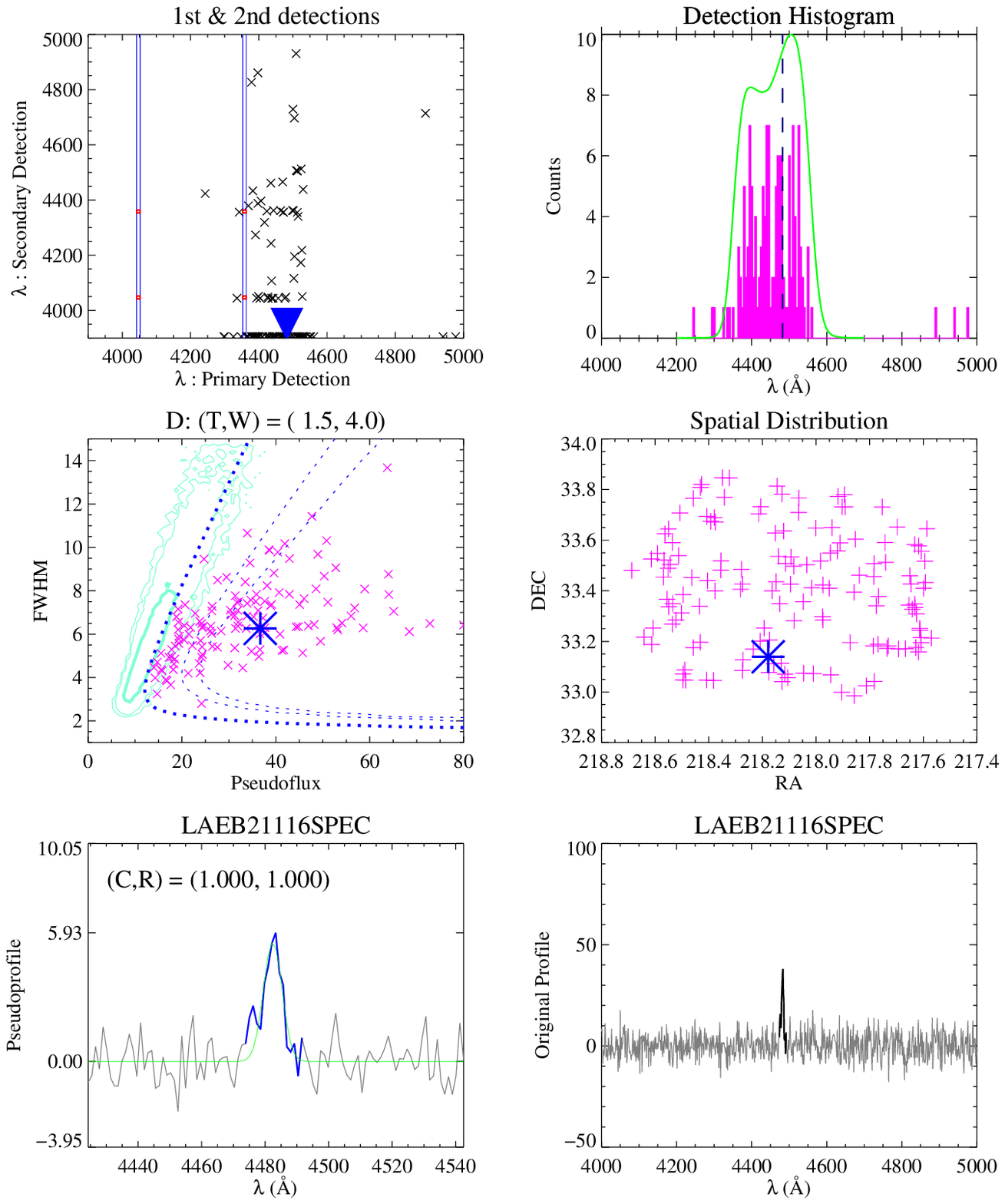}
\caption{An automated customized detection after applying the selection procedures for LAEB21116. We  
observe the differences in the top-left, top-right, and middle-left panels from Figure~\ref{fig:eleven} after the selection procedures.    
An inverted triangle is used instead of a blue asterisk in the top-left panel due to no second detection for LAEB21116; i.e., D100 detection. 
This is a typical reliable detection. 
}\label{fig:twelve}
\end{figure*}

\subsubsection{Customized detections}


In this section, we describe the second stage selection that results in reliable detections.
The bottom panel of Figure~\ref{fig:five} shows the R2 contours, completeness contours, and R1 distribution. 
Since the true reliabilities lie between R1 (conservative false detections) and R2 (generous false detections), 
we need to set a locus excluding R1 fully  and R2 partially. 
Apparently, the contour of  the completeness = 0.5 seems to 
work for this constraint to exclude R1 fully  and R2 partially. 
There is no correct choice; one can choose more generous or strict reliability depending 
on how the detections will be used. Alternatively, all detections may be used, if weighting by the reliability. 
In our method, we use this completeness = 0.5 line to remove low reliability detections. 

To exclude the sky residuals, we reject all detections within $\pm 5$\AA~ from the sky lines, \iion{Hg}{i}  4047\AA~ and 4358\AA. 
If the first detection is excluded by these windows, the second is chosen as the primary detection.
We show two examples, LAEA5569 and LAEB21116, 
to illustrate the process of identifying robust detections. 
In LAEA5569 (shown in Figure~\ref{fig:eleven}), both the first and 
second detections fall at wavelengths affected by systematic errors due to the subtraction of strong sky lines. 
While the detections (just barely) lie within our chosen reliability criterion, 
they are excluded by their wavelength position. In LAEB21116 (shown in Figure~\ref{fig:twelve}), 
the emission line is well detected, highly reliable, and well separated from any region affected 
by sky subtraction systematics. We refer the reader to the figure captions for detailed 
descriptions of the characteristic plots produced by our code.



\begin{figure*}
\centering
\includegraphics[height=3.5 in]{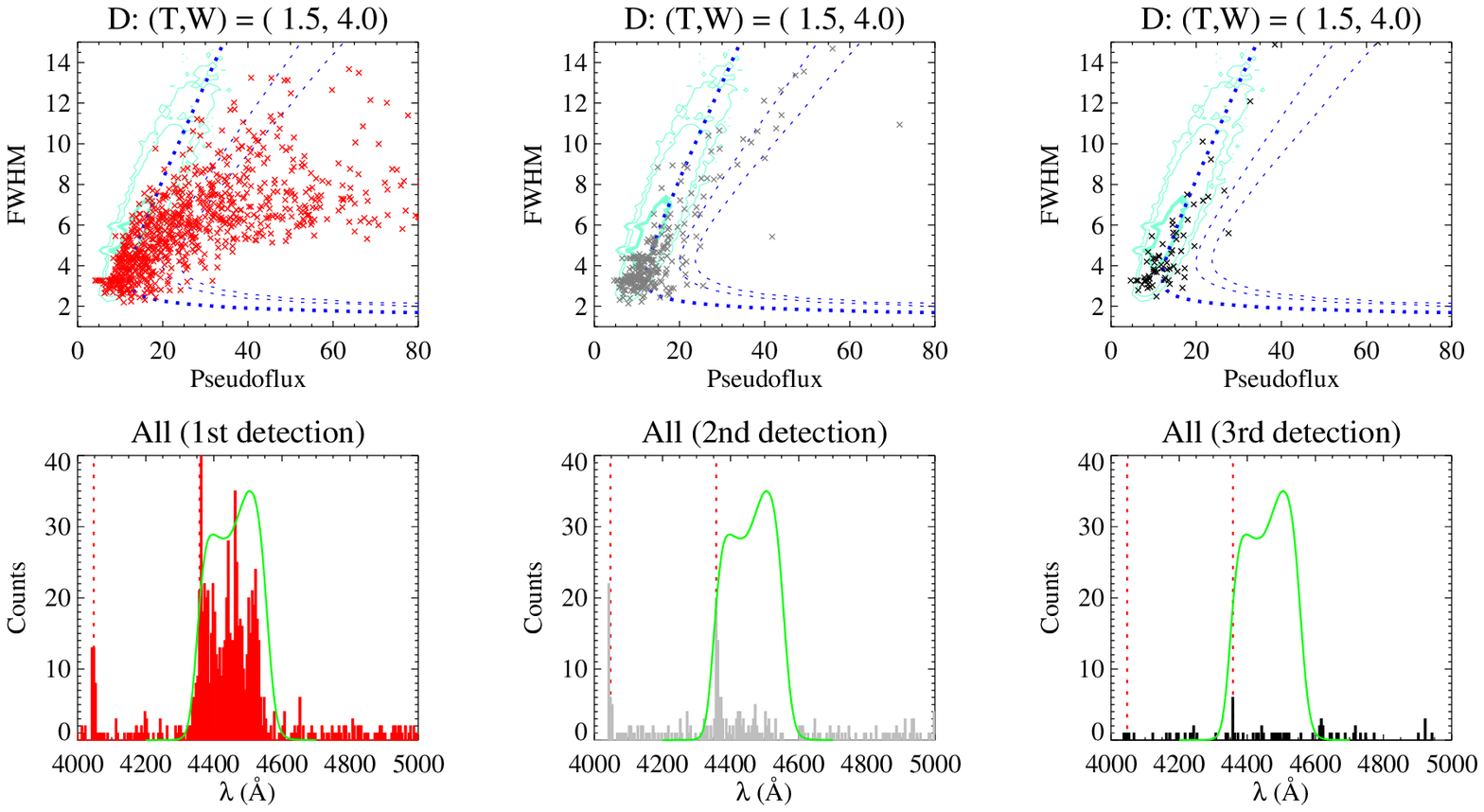}
\caption{The first, second, and third detections from 1574 spectra in all configurations. 
The number of each detection category is 616 for D100, 196 for D110, and 69 for D111, 
totaling 881 having raw detections out of the 1574 observed candidates. 
Like Figure~\ref{fig:nine}, most of the second and third detections are sky residuals or noise detections. 
}\label{fig:thirteen}
\end{figure*} 

\begin{figure*}
\centering
\includegraphics[height=7.5 in]{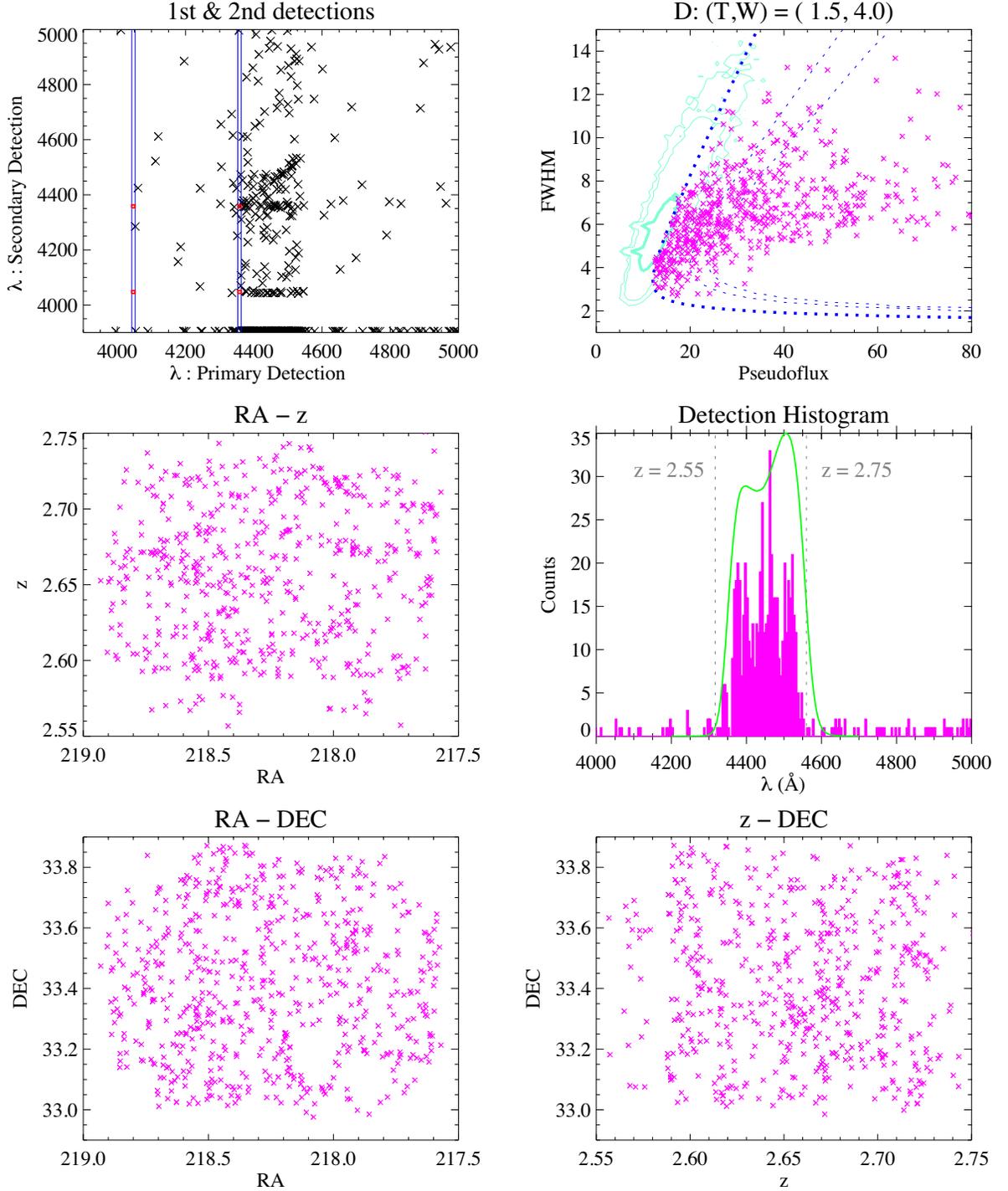}
\caption{The final results of 649 customized detections. 
From 881 raw detections, we obtain 649 customized detections. 
In the top-left panel, we can find the three strips of clustered points. 
The two horizontal strips at 4047\AA~ and 4358\AA~ are produced by the sky residuals 
as the second detections. The third diagonal feature is the ascending strip from 4350\AA~ to 4550\AA,  
where the centroids of the first and second detections are similar, $\lambda_{primary} \approx \lambda_{secondary}$. 
This feature results from the detection of double-peaked Ly$\alpha$ emission lines.
The left-middle, left-bottom, and right-bottom panels show the RA -- DEC -- z distribution.
}\label{fig:fourteen}
\end{figure*} 

\begin{figure*}
\centering
\includegraphics[height=7.5 in]{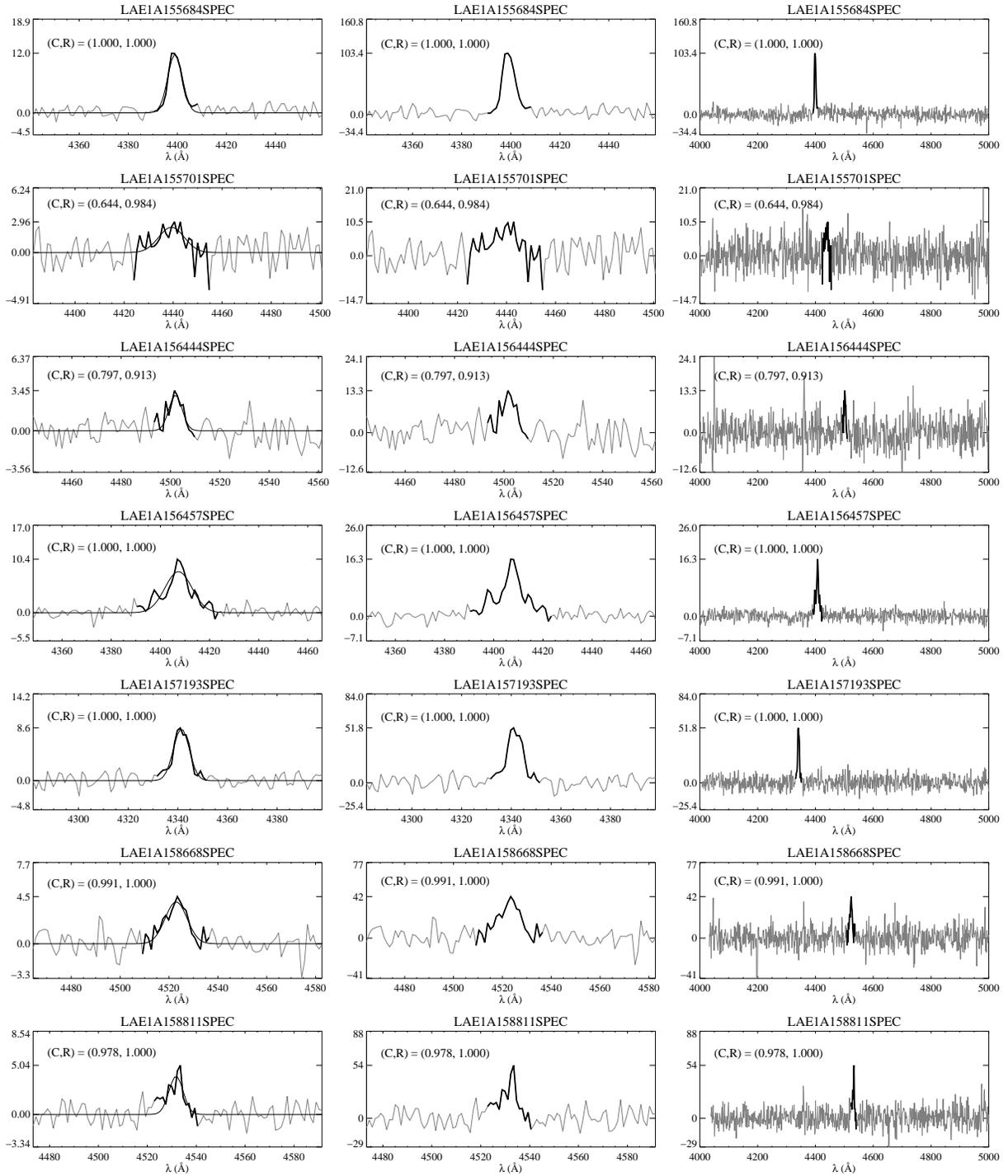}
\caption{The profiles of the final customized detections with C and R values; pseudoprofile (left panel), original profile(middle panel), 
and original profile in the full spectral searching range (right panel). 
The detections of (C,R)$~\approx (1.0,1.0)$ are consistent and clear in manual eye inspections too. 
They are also robust to most fluctuations of baseline due to their high SNR.   
For the detections of low reliability and completeness, the signal features are marginal. 
}\label{fig:fifteen}
\end{figure*} 

\begin{figure*}
\centering
\includegraphics[height=7.5 in]{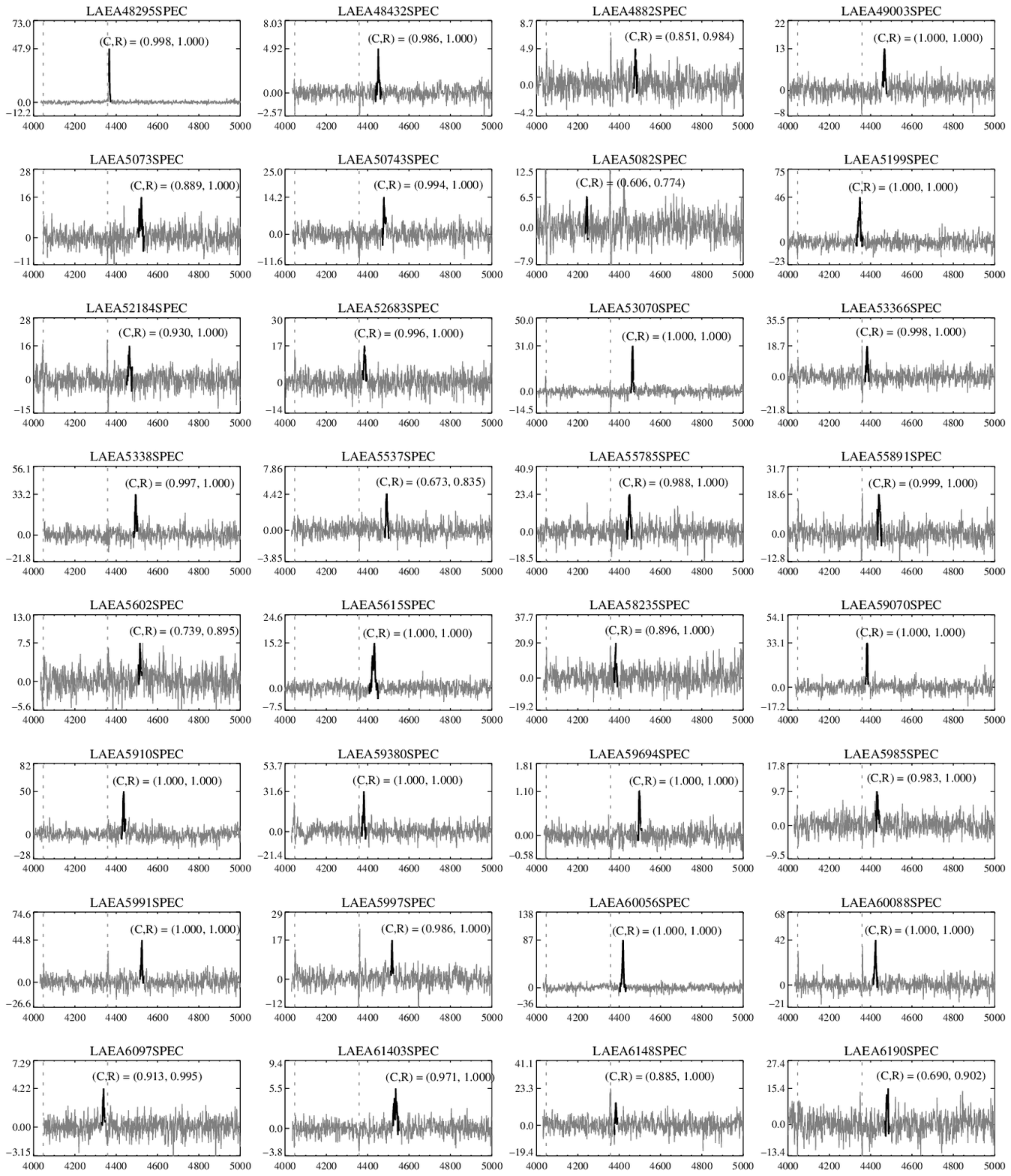}
\caption{Full spectral searching range (as in the third column of Figure~\ref{fig:fifteen}) for 32 additional detections.
The two grey--dotted vertical lines are \iion{Hg}{i} 4047\AA and 4358\AA. 
LAEA4882, LAEA5082, and LAEA5602 show some fluctuations of baselines, 
possibly due to poor sky subtraction or weak underlying stellar continuum. 
In these cases, some weak detections are disputable as enhanced false detections due to uncertain baselines.
}\label{fig:sixteen}
\end{figure*}

\subsubsection{Application to large samples}

We have demonstrated how our method works for data from a single MMT/Hectospec configuration PM1. 
Now we apply our method to all 7 configurations of the 1574 LAE candidate spectra. 

Figure~\ref{fig:thirteen} shows the first, second, and third detections from the 1574 spectra in the same format of Figure~\ref{fig:nine}. 
There are 881 raw detections in the 1574 spectra, 616 of which are first detections and 196 of which are second detections. 
The raw detection rate of all configurations is 0.56, which is lower than 0.73 of PM1 alone. 
The drop in the success rate is partly due to weather (the observations in 2012 
were taken in poor weather and typically had shorter exposure times) 
and partly because of changes in the candidate selection between runs (see Dey et al. in prep for details).
As in Figure~\ref{fig:nine}, most of the second and third detections are sky residuals or noise detections 
following the low reliability contours. 

Figure~\ref{fig:fourteen}  shows the final 649 customized detections from the 881 raw detections. 
The results shown in the top-left, top-right, and middle-right panels 
are the same with Figure~\ref{fig:twelve}. 
There is, however, one interesting feature only revealed in this large sample. 
In the top-left panel, we can find the three strips of clustered points. 
The two horizontal strips at 4047\AA~and 4358\AA~are produced by 
the sky residuals as second detections. This is of no interest. 
The real interesting feature is the ascending strip from 4350\AA\ to 4550\AA,  
where the centroids of the first and second detections are similar, $\lambda_{primary} \approx \lambda_{secondary}$. 
This feature results from the detection of double-peaked Ly$\alpha$ emission lines.
Although many of the Ly$\alpha$ emitters exhibit single-peaked emission lines 
(at least, at our spectral resolution of 6\AA), 21 detections exhibit double peaked lines, 
where both peaks are independently recorded as significant detections by our algorithm. 
The bottom-left panel shows the R.A. and Decl. positions , the middle-left the R.A. and redshift, and 
the bottom-right panel the Decl. and redshift of the 649 detections. The redshifts are calculated from the observed centroids 
and the wavelength of Ly$\alpha$ emission. We can find clusters and filaments from the redshift distributions. 
The heavily populated LAEs near $z \approx 2.67$ form a wall-like structure rather than a compact cluster.  

Figure~\ref{fig:fifteen} shows the randomly chosen 7 detections with their C and R values within the detected LAEs. 
The first column shows pseudoprofiles, the second original profiles, and the third original profiles in the wider spectral range from 4000 -- 5000\AA. 
Figure~\ref{fig:sixteen} shows the 32 more detections with their C and R values in the same format of the third column of Figure~\ref{fig:fifteen}. 
The high significance detections of $C \approx 1.0$  and $R \approx 1.0$ are quite clear and robust signals.  
For lower C and R detections, they become more marginal in visual inspection.  
Overall the automated detections are consistent with visual inspection; especially for high significance detections. 
For marginal detections, we prefer our automated detections, not because of better reliability 
but because of better consistency and the ability to quantify the reliability of the detection, free from the subjectivity of visual inspection.



\section{Summary}

We have presented an automated and quantitative method to detect a narrow emission line within a target spectral range. 
The key point of the method is to generate reliability and completeness contours using Monte Carlo simulations. 
By comparing the contours and the observed detections, 
we can assign reliability and completeness values for each detected feature and 
finalize the detection using customized selection criteria. 

We have applied our method to MMT/Hectospec observations of a sample of candidate Ly$\alpha$ emitters. 
All high significance detections of $(C, R) \approx (1.0, 1.0)$  
are clear and robust in visual inspection. 
For marginal $(C, R)$ detections, 
we prefer our automated detection to visual inspection, 
since the automated method is free from the subjectivity of visual inspection. 
Though the application presented 
here is designed to detect single faint emission lines, 
we can apply our method to more general problems, if we can subtract 
continuum baseline properly. 
One of the challenges of our methodology is that it relies on accurate estimation and subtraction of the continuum, 
which can be a challenging task for data from multifiber spectrographs.
For most cases, we need templates of continuum models and find the optimized one using $\chi^2$ minimization 
(e.g. Tremonti et al. 2004, Bolton et al., Lee et al.). 
Since each kind of object has its own continuum model 
(synthesized stellar spectra for galaxies, quasar templates for quasars, stellar spectral types for individual stars, dust models for infrared emissions), 
the continuum subtraction is better to be a separate process developed independently.  
When combined with a proper method of continuum subtraction, our method can be applied beyond faint emitter detections.


\acknowledgments
We are grateful to an anonymous referee for comments that have improved this 
paper. Observations reported here were obtained at the MMT Observatory, 
a joint facility of the Smithsonian Institution and the University of Arizona. 
The telescope time was granted both by NOAO, through the Telescope 
System Instrumentation Program (TSIP; funded by NSF) and through Steward Observatory.
This paper uses data products produced by the OIR Telescope Data Center, 
supported by the Smithsonian Astrophysical Observatory. 
We are grateful to both the NOAO and Steward Observatory time allocation committees 
for their continuing support of this program. The research activities of SH and AD are supported by NOAO, 
which is operated by the Association of Universities for Research in Astronomy under 
a cooperative agreement with the US National Science Foundation. 
AD's research is also supported in part by the Radcliffe Institute 
for Advanced Study at Harvard University. SH is supported by NASA grants XXXXXXX. 
MKMP is supported by a Dark Cosmology Centre Postdoctoral Fellowship; 
The Dark Cosmology Centre is funded by the Danish National Research Foundation.


\end{document}